\documentclass[]{aastex631}

\newcommand\cd{cd$^{-1}$\,}

\begin{document}

\title{$\beta$ Cephei pulsators in eclipsing binaries observed with {\it TESS}}

\correspondingauthor{Christian I. Eze}
\email{cheze@camk.edu.pl}

\author[0000-0003-3119-0399]{Christian I. Eze}
\affiliation{Nicolaus Copernicus Astronomical Center of the \\
Polish Academy of Sciences, Warsaw, Poland}
\affiliation{Department of Physics and Astronomy,\\
University of Nigeria, Nsukka, Nigeria}

\author[0000-0001-7756-1568]{Gerald Handler}
\affiliation{Nicolaus Copernicus Astronomical Center of the \\
Polish Academy of Sciences, Warsaw, Poland}







\begin{abstract}

The combined strength of asteroseismology and empirical stellar basic parameter determinations for in-depth asteroseismic analysis of massive pulsators in eclipsing binaries shows great potential for treating the challenging and mysterious discrepancies between observations and models of stellar structure and evolution of massive stars. This paper compiles a comprehensive list of massive pulsators in eclipsing binary systems observed with {\it TESS}. The {\it TESS} light curves and Discrete Fourier Transforms (DFT) of a sample of 8055 stars of spectral type B0--B3 were examined for eclipses and stellar pulsations and the ephemerides of the resulting sub-sample of massive pulsators in eclipsing binaries were computed. This sub-sample was also cross-matched with existing catalogues of massive pulsators. Until now, fewer than 30 $\beta$ Cep pulsators in eclipsing binaries have been reported in the literature. Here we announce a total of 78 pulsators of the $\beta$ Cephei type in eclipsing binaries, 59 of which are new discoveries. Forty-three are recognized as definite and 35 are candidate pulsators. Our sample of pulsating massive stars in eclipsing binaries allows for future asteroseismic modelling to better understand the internal mixing profile and to resolve the mass discrepancy in massive stars. We have already started follow-up of some of the most interesting candidates.

\end{abstract}



\section{Introduction} \label{sec:intro}

Asteroseismology is the study or science of determining the interior structure of stars from their oscillations/pulsations \citep[e.g.,][]{Gough1985,Handler2013, Aerts2021}. These oscillations exist in the form of radial or nonradial oscillation modes. Radial oscillation modes are seen in classical pulsators (e.g. Cepheid, RR Lyr stars and Miras) and show strong relevance in observational cosmology  \citep[e.g.,][]{AndersonandRiess2018}. On the other hand, nonradial oscillation modes are of interest in asteroseismology. They are solutions to equations of motion of a star that gets perturbed from its equilibrium \citep{Aerts2021}. The nonradial modes are mainly of two types, depending on which of the forces (pressure or buoyancy) is dominant in restoring the equilibrium. Modes that are restored by the pressure force are called pressure modes (p modes) and those that are restored by buoyancy are termed  gravity modes (g modes). While p modes have large amplitudes in the envelopes of the stars and are characterised by dominant radial motions, g modes have large amplitudes in the deep interior of the star and are characterised by dominant horizontal motions. Another class of nonradial oscillation mode are called mixed modes, that have p-mode character in the envelope and g-mode character in the deep interior and shows excellent probing power in the entire star.

Massive stars also show pulsations \citep[e.g.,][]{Bowman2020} and offer unique opportunities to constrain their properties via asteroseismology \citep{Aertsetal2010}. Three classes of massive pulsators exist, namely: $\beta$ Cephei stars which are early B type main sequence stars with masses of approximately 9 – 17 M$_{\sun}$, low radial order pressure (p), gravity (g) and mixed modes, pulsation amplitudes up to a few tenths of magnitudes and pulsation periods of several hours (approximately 2 to 6 hours) \citep[e.g.,][]{1993SSRv...62...95S,2003SSRv..105..453A,StankovandHandler2005, Handler2013}, 
Slowly pulsating B (SPB) stars which are mid-to-late B stars of masses 3 – 9 M$_{\sun}$ with high radial order g-modes and observationally challenging periods of the order of days (approximately 0.5 to 4 days) \citep[e.g.,][]{1991A&A...246..453W, 1998A&A...330..215W}
and a third group which shows stochastic low-frequency (SLF) variability and have quasi-periodic and time-dependent variability spanning a broad range of periods from order of minutes to several days \citep[e.g.,][]{Bowmanetal2019a, Bowmanetal2019b, Bowmanetal2020}.
Following the relative abundance of pulsators of spectral type B compared to O stars, the majority of the constraints on  the stellar structure and evolutionary theory of massive stars currently are believed to come from {\ensuremath{\beta}} Cep and SPB stars \citep{Aertsetal2019, Bowmanetal2020}. {\ensuremath{\beta}} Cep stars are adjudged to be supernova progenitors and their roles in shaping galaxy dynamics or its chemical enrichment are crucial. Although they predominantly pulsate in p modes \citep{StankovandHandler2005}, they also have g and mixed pulsation modes \citep[e.g.,][]{2017MNRAS.464.2249H}. These make them very important in seismically probing the entire star.  Their understanding gives us insight into the properties of massive stars such as rotational mixing, convective core overshooting e.t.c. as well as the transition from main sequence to hydrogen-shell burning evolution of massive stars \citep{NeilsonandIgnace2015}. Owing to these features, especially core convection, uncertainties in the stellar structure and evolution theory are largest for stars of O and B categories \citep{Pedersenetal2019}, and hence, compound the problem or mystery of mass discrepancy in massive stars. The mass discrepancy problem is a situation where the masses of stars inferred from spectroscopy are different from the masses derived from models of stellar evolution \citep{Herreroetal1992, Tkachenkoetal2020}.

Massive stars are predominantly found in multiple systems \citep{Sanaetal2012, Sanaetal2014, Kobulnickyetal2014, SouthworthandBowman2022}, hence, a substantial fraction are expected to also be located in eclipsing binaries. \citet{Southworth2012} in line with \citet{Russell1948} and \citet{Batten2005} described ‘eclipse’ as the royal road to stellar astrophysics. By modeling eclipses,  model-independent precise stellar parameters (e.g. radius, mass), which serve as invaluable calibrators for stellar evolution theory \citep{Torresetal2010, Pedersenetal2019} are obtained. Dynamical masses deduced from eclipsing binary modelling  and asteroseismic masses offer unique opportunities to constrain the physics of stellar evolution models  to treat the challenging and mysterious discrepancies between observation and models \citep{Tkachenkoetal2020}. Unfortunately, the absolute numbers of reported massive stars in eclipsing binary or multiple system are significantly smaller compared to their low mass counterparts \citep{Kirketal2016, Pedersenetal2019}.  However, {\it TESS} \citep{Rickeretal2015} is out to change this narrative. It is the first precision photometry mission that surveys (almost) the whole sky and has released, in large amounts, time resolved photometry of stars of O and B spectral types \citep[e.g.,][]{Handleretal2019} with a precision comparable to {\it Kepler} and K2 but for stars five magnitudes brighter owing to its smaller telescope aperture. Since the advent of the {\it TESS} mission, there has been concerted effort by researchers in the field to obtain comprehensive and adequate sample size of massive pulsators for massive star asteroseismology to improve  the physics of the stellar structure and evolution models of massive stars using the {\it TESS} data. These  efforts are to compensate for the lack of sufficient number of such stars observed with high-precision space photometry prior to {\it TESS}  \citep{Aertsetal2010, Handler2013} and to gather a pool of suitable candidates for in depth asteroseismic analysis. This major barrier of small sample size thus far hampered the general asteroseismic understanding of high mass stars.  

A number of authors have published catalogues of massive stars in the recent past. \citet{Pedersenetal2019} presented the classification of variability of 154 massive stars with spectra types O and B that were observed by {\it TESS} in short cadence (2 minutes). Their sample consisted of  single or binary stars of diverse variability and was aimed at establishing an unbiased sample for O and B stars for future asteroseismology. A similar search has been conducted by \citet{Burssensetal2020}, who combined the TESS photometry with spectroscopy allowing the stars to be more accurately placed on the HR diagram. Forward asteroseismic modelling of single $\beta$ Cep stars has also been conducted by \citet{Burssensetal2023}.

To combine the strengths of eclipsing binary modelling  and asteroseismology, a lot of other recent searches have been either narrowed down to massive pulsators in eclipsing binary systems or done in larger scales to capture more massive eclipsing binaries in the sample \citep{Southworth2015,  Southworthetal2020, Southworthetal2021,  Ijspeertetal2021, Zarietal2021, SouthworthandBowman2022}. While some of these authors focused mainly on identifying eclipsing binaries, these works still play a significant role in the search for {\ensuremath{\beta}} Cep pulsators in eclipsing binaries as they provide lists of already existing eclipsing binaries.  Whereas \citet{Southworth2015}, \citet{Southworthetal2020, Southworthetal2021} and \citet{SouthworthandBowman2022} have small numbers of massive pulsators in their samples as they considered heterogeneous sample of pulsating eclipsing binaries with  predominantly short orbital period ($<$27 d) owing to having 1+ TESS sectors, \citet{Ijspeertetal2021}, on the other hand, published a large sample of massive stars comprising single stars and eclipsing binaries in their effort to identify eclipsing binaries in OBA-type stars. However, they limited their search to stars of {\it TESS} magnitude below 15 with colour indices $J-H < 0.045$ and $J-K < 0.06$. There is a large overlap between the catalogue of OBA-type stars  compiled by  \citet{Ijspeertetal2021} and that compiled by  \citet{Zarietal2021} which used a Gaia magnitude cut of $G < 16$ as one of their major selection criteria. Here, we  aim to compile a comprehensive catalogue of early B-type (B0--B3) pulsators in eclipsing binaries  observed by {\it TESS} with a particular focus on {\ensuremath{\beta}} Cep stars, in order to harness the combined potentials of eclipsing binary stars and asteroseismology to probe the evolution and properties of massive stars. The spectral range we selected is expected to yield a comprehensive homogeneous sample of massive main sequence pulsators with self-excited p, g and mixed modes needed for the overall asteroseismic probe of massive pulsators.  In section 2, we describe the sample, its observation and selection criteria. In section 3, we describe the analysis. We discuss the results of the  variability classification and periodicity  of the pulsators in our sample in section 4 and draw the necessary conclusions in section 5. 

\begin{figure}[ht!]
\plotone{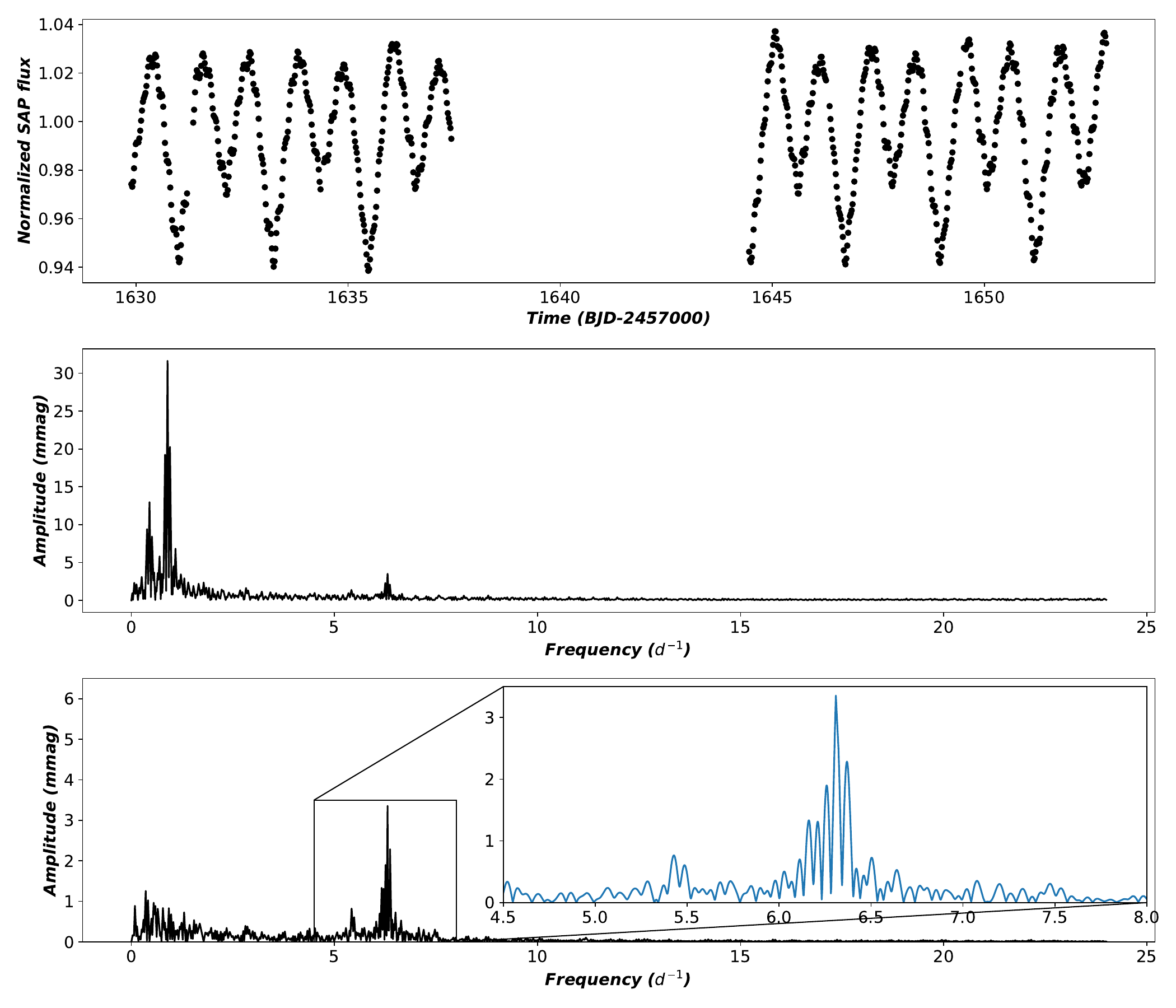}
\caption{An example figure of the light curve and DFT of a massive  pulsator in eclipsing binaries (HD 329379) in our sample. The upper panel shows the TESS light curve. The middle panel shows the DFT of the light curve shown in the upper panel whereas the lower panel shows the DFT of the residual light curve after removing the orbital light variations. The region showing the dominant $\beta$ Cep pulsations is also zoomed into and shown in the inset in the lower panel.}
\label{fig:sample}
\end{figure}

\newpage
\section{Sample Selection and Observation} \label{sec:style}

The primary photometric data used for this work are the {\it TESS} 30-minute cadence light curves (QLP data) \citep{2020RNAAS...4..204H,2020RNAAS...4..206H}, obtained from the Barbara A. Mikulski Archive for Space Telescopes (MAST) and observed in the first  two cycles of {\it TESS}, covering sectors $1-26$.  The QLP light curves of 8055 stars of spectral type B0--B3 were analysed. The list of the stars was first obtained from a compilation of published spectral type catalogues done by Luis Balona (private communication; see also arXiv:2212.10776) and the {\it TESS} light curves downloaded from MAST. For stars in our sample with available 2 minute cadence data at the time of this work, we also examined the 2 minute cadence light curves. The sample spans ranges of $3.23-12.33$ in V magnitude and $3.01-12.96$ in B magnitude, respectively. As a result, it includes both bright and faint stars with varying morphology. In terms of morphology, eclipsing binary stars are classified as detached, semi-detached and over-contact binaries \citep[see][for details]{KallrathandMilone2009}. While detached eclipsing binaries are more convenient to work with, comparing an ensemble of detached binaries with those experiencing strong tidal forces, one understands the importance and role of tidal forces in the internal processes \citep{Ijspeertetal2021}. 
Hence, our search is not restricted to detached eclipsing binaries but includes massive {\ensuremath{\beta}} Cep pulsators in eclipsing binaries of different {\it morphological classifications} with self-excited pulsations. Irrespective of the importance of Stochastically Low Frequency (SLF) variables in the study of massive stars via internal gravity waves, we ignored them during the selection of pulsators from our sample.
We also ignored the SPBs in our sample during the selection owing to their mass range which is not within the {\ensuremath{\beta}} Cep regime of interest. However, any hybrid massive pulsator with self-excited pulsations and {\ensuremath{\beta}} Cep component was included in the sample of pulsators. Hereinafter, we shall refer to the entire list of the stars examined as "the sample" and refer to the list of eclipsing binaries in the sample with {\ensuremath{\beta}} Cep pulsating component(s) as "the sample of pulsators".

\section{Analysis} \label{sec:floats}
To preselect candidates, we first plotted the light curves and the Discrete Fourier Transforms (DFT) of each of the 8055 stars. Via visual examination of the plots, stars which show eclipses as well as frequency peaks in the regimes of the $\beta$ Cephei were preselected. Here, we considered pulsations with frequencies, $f \geq 3$ d$^{-1}$ as {\ensuremath{\beta}} Cep pulsations.  Owing to the effect of binarity, some of the frequency peaks adjudged to be independent  frequencies sometimes appear to be harmonics of the orbital frequency or could be hidden within a forest of peaks in the DFT caused by the orbital light variations. In a few cases, the pulsational signals were of higher amplitude than the eclipses, which would then manifest themselves as a series of low-frequency harmonics in the DFT. As a result, successive prewhitening of each of the light curves of the stars in the preselected sample was done using the {\it Period04} software \citep{LenzandBreger2005}. 

The {\it Period04} software \citep{LenzandBreger2005} applies discrete frequency Fourier analysis and simultaneous multi-frequency least square fitting. Calculation of light curve fits for multi-periodic signals such as harmonics, combination and equally spaced frequencies are also possible in the program. To remove the effect of binarity from the light curve, we fitted the orbital frequency and all its detected harmonics and subtracted the fit from the light curve resulting in a residual that would contain the pulsational signals only. We further prewhitened this residual to check for independent frequency peaks. For the prewhitening, we used an upper frequency of 23.5 d$^{-1}$, which is slightly below the Nyquist frequency and accepted independent pulsation frequencies with S/N $\geq$ 4.6 as real frequencies \citep[cf.][]{2021AcA....71..113B}. For the sake of this classification, we did not do exhaustive prewhitening of the residual to extract all the frequencies. Since our interest is to compile massive pulsators and not the exact number of pulsations they have, we only considered at least one to five dominant peaks and in cases where hybrid pulsation is suspected, we continued the prewhitening until the given object could be established or disproved as a hybrid pulsator.
For frequencies between 2 and 4 $\rm d^{-1}$, it is difficult to classify the pulsators as $\beta$ Cep, SPB or hybrid using only the frequency range owing to the fact that rotation can split multiplets of non-radial p and g modes into the regimes of each other. Combination frequencies can also fall into this frequency domain. The latter have been checked for and are not used to classify the variability. A small number of stars do show signals between 2 and 4 $\rm d^{-1}$, but determining their exact cause is outside the scope of this work. Figure 1 shows an example light curve and DFT of a hybrid pulsator in the sample.

The stars in the preselected sample, which have shown independent pulsation peaks were subjected to further checks for false eclipses and blends. The light curves of the stars (especially in crowded regions on the sky) are sometimes modulated by stray light from neighbouring stars which blends with their light. Owing to blends, there have been incidences of false eclipses in stars or pulsations wrongly attributed to stars that they do not originally come from. To correct for blends, we conducted photometric checks and analyses using Eleanor v2.0.5 \citep{Feinsteinetal2019, Brasseuretal2019, Burkeetal2020}. Eleanor is an open source python based package used for downloading, extraction, analysis and visualization of flux corrected light curves from {\it TESS} Full Frame Images (FFIs). It is also used for cross-matching of light curves. It takes, as input, a TESS Input catalogue ID, coordinates (RA, DEC) or a Gaia source ID of a star observed with TESS and returns, as a single object, a light curve and an accompanying target pixel data. It allows one to plot and visualise pixel-to-pixel light curves and periodograms and possibly match the flux corrected light curves with the existing TESS light curves or overlay the target with the Gaia sources in the field.  
Eleanor also uses TESS-Point  \citep{Burkeetal2020}, which is a high precision TESS pointing tool that converts coordinates given in RA and DEC to the TESS detector pixel coordinates and TESScut \citep{Brasseuretal2019}, which is a tool that makes a cutout, for a given region of the sky, of a TESS FFI time series under the hood to extract its data products. Using Eleanor, we extracted the {\it TESS} target pixel files, plotted colour-coded pixel-to-pixel flux corrected light curve and DFT of the target(s) where blends are suspected. For targets where the Eleanor default aperture appeared not to be very suitable, we defined a custom aperture mask by visually examining the pixel data and defining an optimal aperture size that captures most of the light from the target without reducing the size to the extent that could compromise the quality of the light curves that would be extracted. Figure 2 shows the cut out session of the aperture and pixel-to-pixel light curves of one of our targets (V1216 Sco). Whereas Eleanor light curves could be used independently for our further detailed analysis, we used it only as a photometric checking tool, for targets that could be blended, to ascertain whether or not the signatures of eclipses and pulsations observed in the QLP light curves are also present in the Eleanor data products, where blends and systematic effects are already taken into account. To rule out the effect of contamination by other possible sources within the aperture or pixel, we over-plotted the optimal apertures over Gaia sources of 5 mag fainter than the targets. 
According to the magnitude-flux relation defined as $m_{i}$$-$$m_{j}$ $=-$2.5\,log$_{10}$($b_{i}$/$b_{j}$) where $m_{i}$ $\&$ $m_{j}$ are apparent magnitudes, and $b_{i}$ $\&$ $b_{j}$ are fluxes,  we thus assumed that a star which is 100 times fainter than the target has negligible blend or modulation effect on the light curves and pulsations of the target(s). By examining the pixel-to-pixel light curves and the overlaid plots, we rule out possible effects of blends  and either accept or reject the pre-identified eclipses and pulsations as genuine. On the other hand, independent pulsations could be matched directly with the stars within the {\it TESS} pixel using another python module called {\it TESS}-Localize \citep{HigginsandBell2022}. In this case, individual pulsations are traced to their origin within the {\it TESS} target pixel file. A signal with a periodogram peak that exceeds 7.5 times the average local noise level in the periodogram can be localized. This latter approach was only considered in situations where it becomes ambiguous to decipher, from the former, where the pulsation(s) come from. As pointed out earlier, we only considered a few dominant peaks which were used for the purpose (where necessary). In cases where the peaks were traced to neighbouring stars, more pulsation  frequencies of the star were extracted and subjected to similar procedure using {\it TESS}-Localize until the target is established or disproved to have pulsations. 

\begin{figure}[ht!]
\plotone{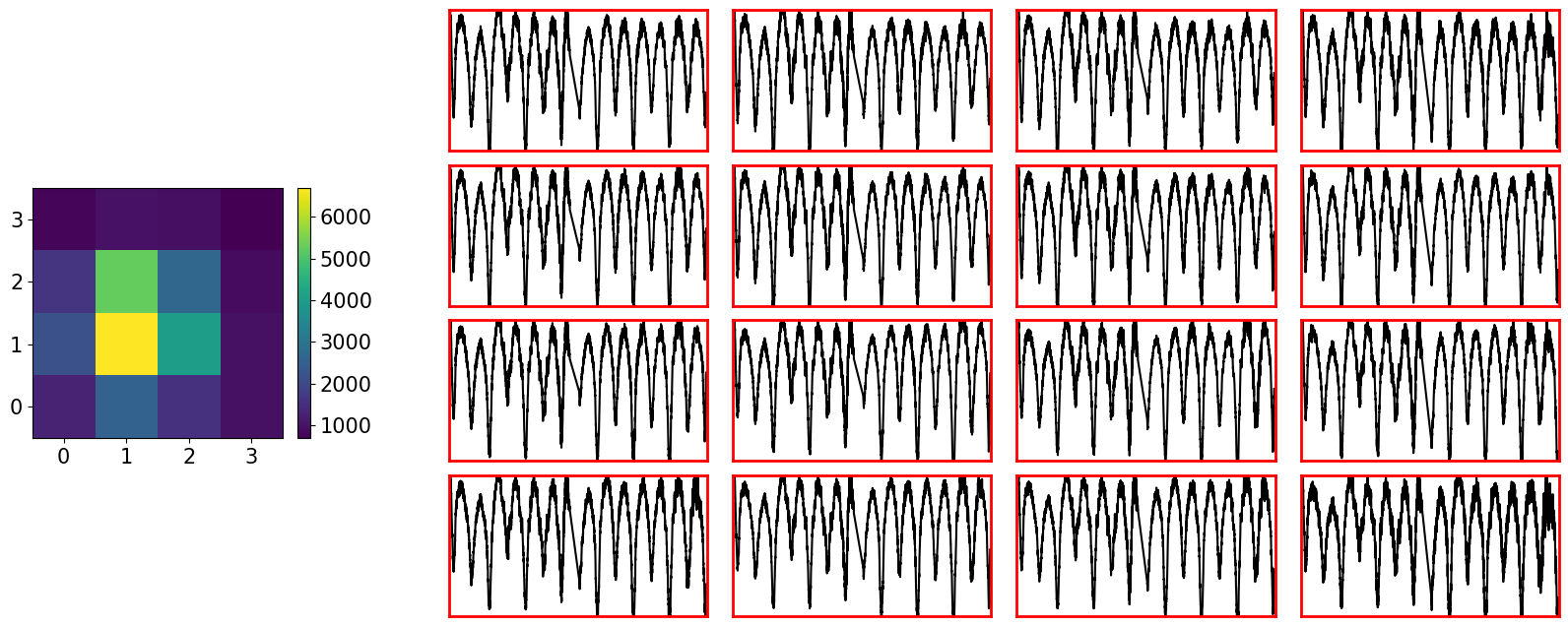}
\caption{A cut-out session of the pixel-to-pixel plot of the aperture mask for V1216 Sco. 
\label{fig:cutout}}
\end{figure}

To ensure that the stars in the sample of pulsators are within the mass range suitable for the observed photometric variability, we placed them in the HR diagram shown in Figure 3, using the information available in the literature. We plotted the luminosity, L as the ordinate and effective temperature, $T_{\rm eff}$ as the abscissa. The luminosity is computed using the equation, log$(L/L_\odot) = 0.4\times(4.74-M-BC)$, where $M = m-$5$\times$log(1000/$\pi$)$+$5$-$$A_0$, $BC$ is the bolometric correction, $m$ is the apparent magnitude in the V band, $\pi$ is the parallax and $A_0$ is the interstellar extinction. $\pi$ is obtained from the Gaia DR3 catalogue \citep{Gaiacollaborationetal2016b,Gaiacollaborationetal2023j}. The $T_{\rm eff}$ is the effective temperature from Gaia DR3 using the hot star pipeline and obtained together with $A_0$ from this source. The bolometric correction is calculated in line with \cite{1996ApJ...469..355F}. The theoretical evolutionary tracks are also computed for the masses 4, 5, 6, 8, 10, 12, 15, 20 and 25$M_\odot$ with the Warsaw-New Jersey evolution and pulsation code (described, for instance, by \citealt{1998A&A...333..141P}) using $Z = 0.012$, $X = 0.700$ and $v \sin i_{ZAMS}=100\,kms^{-1}$. The theoretical instability strips are assigned using the same conditions \citep{1999AcA....49..119P}. We see relatively good agreement of the positions of the stars in this diagram with the $\beta$ Cep instability strip, but with the caveats that the external uncertainties of the Gaia DR3 hot star effective temperatures are not well known and that our targets are binary systems, i.e. the measured temperature is a weighted average of the temperatures of both components.

\begin{figure}[ht!]
\plotone{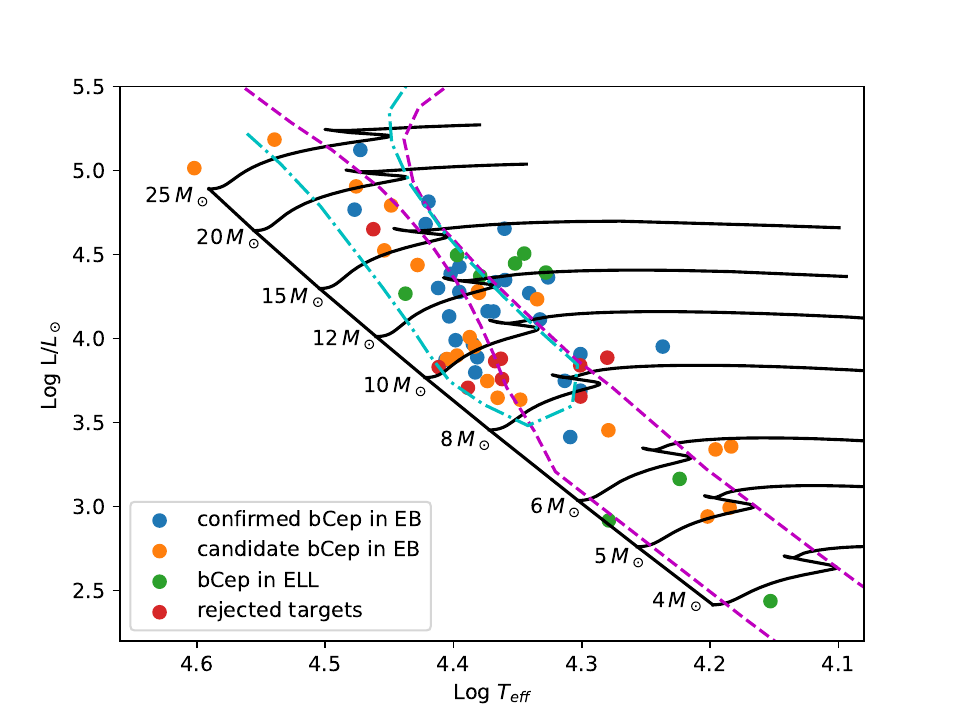}
\caption{HR diagram of the stars listed in Tables 1-4. The dashed magenta line denotes the SPB instability strip whereas the dashed-dotted cyan line denotes the $\beta$ Cep instability strip.  Stars with missing parameters (e.g. $A_0$, $T_{\rm eff}$) are not captured in the HR diagram. 
\label{fig:HR}}
\end{figure}

To compute the ephemerides of the sample of pulsating stars, period analysis was conducted. We computed the preliminary  orbital periods of the systems using the box-fitting least squares, BLS \citep{Kovacsetal2002} and Lomb-Scargle, LS \citep{Scargle1982} periodograms incorporated in the  LC periodogram solver of PHOEBE 2 program \citep{Prsaetal2016}.  The precision of the period obtained with PHOEBE Estimator may improve during the detailed  modelling of the light  curves of the individual stars by applying PHOEBE Optimizer and emcee, which is beyond the scope of this paper. Consequently, the estimated periods obtained from PHOEBE  or available in literature were used as input priors for fitting of the orbital harmonic model and further refined during prewhitening using Period04. Harmonics that have amplitudes that satisfy the same S/N criteria as the pulsations were used to build the orbital harmonic binary model. The period reported in Tables 1--4 for the stars  is the final period obtained in the Period04 program and which fits the orbital binary model the best.

\section{Results and Discussion} \label{sec:floats}
\subsection{Classification of pulsators} \label{subsec:tables}
\begin{table}[ht]
\caption{List of definite {\ensuremath{\beta}} Cep pulsators in eclipsing binary systems with their orbital ephemerides and dominant pulsation frequencies. P is the period, T is the epoch given in BTJD, F is the dominant $\beta$ Cep pulsation, A is the amplitude of the dominant $\beta$ Cep pulsation and SpType is the spectral type of the star. The dominant $\beta$ Cep pulsations and their amplitudes are given with their analytical errors.}  
\begin{center}
\resizebox{\textwidth}{!}{\begin{tabular}{cccccccccccccc}
    \hline
       Name 	&	TIC ID	&	      RA                     & Dec      	&	Variability 	&	  Bmag  	&	  Vmag  	&	P (d)	&	T (BTJD)	&	F (d$^{-1}$)	&	A (mmag)	&	S/N	&	SpType   	&	Ref \\	
\hline																									
16 Lac       	&	129538133	&	22 56 23.63 & +41 36 13.95	&	EB+bCep	&	5.436	&	5.587	&	12.0951(5)	&	1748.220(2)	&	5.9112(2)	&	6.19(6)	&	127.9	&	B2IV           	&	L10, J80, W51	\\
$\lambda$ Sco 	&	465088681	&	17 33 36.52 & -37 06 13.76	&	EB+bCep	&	1.49	&	1.63	&	5.949(1)	&	1635.023(30)	&	4.68035(7)	&	6.64(2)	&	49.6	&	B2IV+DA7.9 	&	SWB22a	\\
BD+44 3594     	&	353099086	&	20 49 11.59 & +45 24 39.79      	&	EB+bCep+SPB	&	10.37	&	9.85	&	$>$33	&	2809.36(40)	&	6.8169(9)	&	0.75(7)	&	16.6	&	B1:V:nep       	&		\\
BD+44 3664     	&	330081196	&	20 59 55.99 & +45 20 12.94	&	EB+bCep	&	10.64	&	10.15	&	4.80323(6)	&	1713.45(5)	&	7.4862(1)	&	7.03(4)	&	66.8	&	B1Vn/B9             	&		\\
CD$-$38 4128     	&	134522557	&	08 09 28.64 & -38 51 29.28	&	EB+bCep	&	10.12	&	9.75	&	5.730(2)	&	1494.955(22)	&	5.67991(6)	&	6.47(2)	&	93.7	&	B1V            	&	BO20	\\
CD$-$51 9984     	&	314529804	&	16 18 22.91 & -51 26 24.27	&	EB+bCep	&	10.59	&	10.25	&	9.79258(3)	&	1632(6)	&	5.744(2)	&	5.9(4)	&	11.6	&	B0.5III        	&	BO20	\\
CPD$-$41 7746    	&	339570153	&	16 54 29.49 & -41 39 14.95	&	EB+bCep	&	9.47	&	9.24	&	6.3467(3)	&	1633.57(10)	&	8.3307(2)	&	1.87(2)	&	56.8	&	B0.5V          	&		\\
CPD$-$45 3109    	&	28957011	&	08 49 35.49 & -46 23 19.27	&	EB+bCep	&	9.796	&	9.58	&	2.7139(1)	&	1519.1(2)	&	5.18739(7)	&	11.76(8)	&	56.4	&	B2/4           	&		\\
CZ Vel      	&	355656323	&	09 10 44.46 & -50 42 40.61	&	EB+bCep	&	11.57	&	10.8	&	5.1929(1)	&	1519.35(20)	&	4.5871(5)	&	2.9(1)	&	33.4	&	B2III/B6             	&	DK04	\\
EK Cru      	&	379012185	&	12 02 58.47 & -62 40 19.23	&	EB+bCep+ SPB 	&	8.1	&	8.11	&	4.7447(1)	&	1572.36(10)	&	6.1436(2)	&	0.54(1)	&	41.2	&	B1V            	&	OW05	\\
EO Aur 	&	408937625	&	05 18 21.07 & +36 37 55.36	&	EB+bCep	&	7.9	&	7.83	&	4.06550(3)	&	1830.47762(5)	&	12.1462(5)	&	0.75(1)	&	40.7	&	B0V+B3V 	&	SWB22a	\\
HD 108628      	&	450918869	&	12 29 06.03 & -62 28 00.15	&	EB+bCep	&	10.04	&	9.76	&	4.222(5)	&	1600.47(7)	&	6.7419(5)	&	2.34(5)	&	46.9	&	B2II           	&	PP08	\\
HD 112026      	&	436382800	&	12 54 18.43 & -60 53 38.93	&	EB+bCep	&	8.69	&	8.66	&	 43.205(1)	&	1622.06(20)	&	6.3869(4)	&	2.22(4)	&	46.9	&	B0/1IV         	&		\\
HD 112485      	&	437617380	&	12 57 51.50 & -60 48 56.08	&	EB+bCep	&	9.57	&	9.53	&	5.372(1)	&	1604.95(90)	&	9.439(1)	&	1.16(6)	&	20.3	&	B1/3(III)      	&	IJ21	\\
HD 113742      	&	440817830	&	13 06 53.75 & -61 56 38.01	&	EB+bCep	&	9.53	&	9.2	&	15.68844(8)	&	1600.43(10)	&	6.936(1)	&	0.66(4)	&	10.1	&	B1/2(III)      	&		\\
HD 150927      	&	78636551	&	16 45 44.06 & -38 10 02.80	&	EB+bCep	&	9.68	&	9.42	&	14.33(2) &	1629.95(20)	&	4.84325(2)	&	116.99(9)	&	186	&	B2/3Ib         	&	PP08	\\
HD 151791      	&	246552414	&	16 51 35.18 & -44 29 35.79	&	EB+bCep	&	9.73	&	9.5	&	3.9173(1)	&	1631.97(20)	&	5.227(1)	&	0.68(4)	&	25.2	&	B2/3Ib/II      	&		\\
HD 152268      	&	339680203	&	16 54 16.59 & -40 58 59.25	&	EB+bCep	&	8.97	&	8.98	&	3.4229(5)	&	1630.75(10)	&	10.0996(8)	&	1.19(4)	&	35.9	&	B1/2Ib/II      	&	IJ21	\\
HD 157400      	&	158688754	&	17 24 36.89 & -35 50 19.75	&	EB+bCep	&	10.1	&	9.68	&	8.4527(4)	&	1630.5(2)	&	11.8628(7)	&	1.22(4)	&	20.3	&	B3/5           	&		\\
HD 188891      	&	171502734	&	19 55 44.76 & +40 23 30.25	&	EB+bCep+SPB	&	7.29	&	7.3	&	161.25	&	1731.0(4)	&	4.971(1)	&	0.67(8)	&	5.3	&	B1V            	&	IJ21	\\
HD 227877 	&	91111448	&	20 08 23.07 & +35 27 33.48	&	EB+bCep	&	9.37	&	9.3	&	1.70664(4)	&	1683.68(5)	&	13.2182(5)	&	0.88(4)	&	36.2	&	B1:IV:nn 	&	C22	\\
HD 254346      	&	426520557	&	06 16 57.32 & +22 11 41.96	&	EB+bCep	&	10.13	&	9.74	&	5.4316(1)	&	2478.5(2)	&	9.2399(2)	&	0.256(6)	&	23.6	&	B2:III:        	&	LB20	\\
HD 303115      	&	458263480	&	10 38 33.37 & -59 21 31.65	&	EB+bCep	&	10.57	&	10.3	&	4.618(1)	&	2286.35(50)	&	5.73966(4)	&	8.65(3)	&	181.9	&	OB-/B5             	&		\\
HD 329379      	&	122314621	&	17 01 05.87 & -45 42 04.24	&	EB+bCep+SPB+ELL 	&	10.7	&	9.83	&	2.2464(4)	&	1631.0(3)	&	6.3086(8)	&	3.4(1)	&	53.3	&	B0II           	&		\\
HD 339003      	&	10891640	&	19 51 02.86 & +25 57 15.43	&	EB+bCep	&	10.43	&	9.93	&	6.163(2)	&	1685.1(2)	&	6.7993(3)	&	12.6(2)	&	43.1	&	B0.5III        	&	LB20	\\
HD 344880      	&	451932686	&	19 45 42.31 & +23 59 04.03	&	EB+bCep	&	9.96	&	9.34	&	54.49399(1)	&	1683.6(4)	&	9.4923(8)	&	1.83(7)	&	32.9	&	B0.5III:nn     	&	LB20 	\\
HD 92024       	&	458076434	&	10 36 08.33 & -58 13 04.36	&	EB+bCep	&	8.9	&	9	&	8.3249(8)	&	2285.135(40)	&	7.1635(1)	&	5.59(3)	&	41.4	&	B1III          	&	F05	\\
HD 96355       	&	466528132	&	11 05 26.38 & -61 26 04.20	&	EB+bCep	&	9.87	&	9.74	&	4.360(2)	&	1574.336(20)	&	6.350(2)	&	0.26(2)	&	16.7	&	B0/1(III)      	&	IJ21	\\
HQ CMa 	&	106830354	&	07 20 54.92 & -26 57 49.81	&	EB+bCep/SPB	&	5.816	&	5.986	&	$-$	&	$-$	&	$-$	&	$-$	&	$-$	&	B3V 	&	SWB22a	\\
LR Ara      	&	447530589	&	16 53 37.19 & -61 35 11.27	&	SPB?+bCep+EB	&	10.6	&	10.7	&	1.5195(3)	&	1626.92(22)	&	4.498(2)	&	0.62(6)	&	7.5	&	B2             	&	BD80\\	
LS I +61 145   	&	406965391	&	00 22 26.52 & +61 49 39.75	&	EB+bCep	&	11.3	&	10.9	&	1.84633(5)	&	1767.71(10)	&	7.540(2)	&	0.29(2)	&	9.9	&	B1V            	&	IJ21	\\
LS I +63 36    	&	359042331	&	00 05 00.85 & +63 49 33.21	&	EB+BCep+DSCT?	&	11.52	&	11.06	&	$-$	&	1782(12)	&	4.319(1)	&	1.1(1)	&	10	&	B0V            	&		\\
TYC 3699-160-1 	&	245470639	&	02 35 50.71 & +58 35 20.83	&	EB+bCep+SPB	&	13	&	12.33	&	3.0687(9)	&	1794.42(10)	&	5.819(1)	&	2.0(1)	&	20.8	&	B0             	&		\\
TYC 4050-2830-1	&	458879750	&	02 22 45.39 & +62 25 36.98	&	EB+bCep+SPB? 	&	11.77	&	11.53	&	11.039(6)	&	1799.506(9)	&	9.239(1)	&	0.57(3)	&	13.4	&	B3             	&		\\
V1061 Cen   	&	334443373	&	14 14 56.81 & -61 14 18.42	&	EB+bCep	&	9.71	&	9.56	&	2.2096(3)	&	1600.68(10)	&	10.142(6)	&	0.22(6)	&	9.3	&	B2II/III       	&	OT03	\\
V1166 Cen   	&	443262289	&	13 15 51.32 & -63 53 03.32	&	EB+bCep	&	8.84	&	8.81	&	13.4551(3)	&	1604.35(20)	&	10.155(1)	&	1.17(6)	&	24.2	&	B1/2V          	&	AG12	\\
V1216 Sco   	&	247315421	&	16 54 57.71 & -43 56 27.17	&	EB+bCep	&	10.93	&	10.17	&	3.9213(6)	&	1632.95(15)	&	5.5517(5)	&	10.6(2)	&	38.2	&	B0             	&	OT03	\\
V2107 Cyg 	&	42244951	&	20 08 45.77 & +37 14 13.36	&	EB+bCep/SPB	&	8.73	&	8.63	&	4.2846(4)	&	1712.70759(6)	&	4.2008(3)	&	3.43(9)	&	25.3	&	B1III 	&	SWB22a	\\
V4386 sgr      	&	60433558	&	18 14 42.14 & -33 08 27.22	&	EB+bCep	&	8.3	&	8.44	&	10.798(1)	&	1664.20(20)	&	6.7726(3)	&	7.63(9)	&	78.8	&	B1II           	&	PP08, C22	\\
V453 Cyg 	&	90349611	&	20 06 34.97 & +35 44 26.27	&	EB+bCep	&	8.52	&	8.4	&	3.8900(4)	&	1684.2(1)	&	4.9459(2)	&	3.05(4)	&	47.7	&	B0.4IV+B0.7IV 	&	SW20	\\
V916 Cen	&	322078735	&	11 42 25.35 &  -62 28 37.47	&	EB+bCep	&	9.68	&	9.63	&	1.46322(5)	&	1573.1113(60)	&	4.451427(6)	&	13.2(1)	&	66.7	&	B0.5IVne	&	PP08	\\
VV Ori 	&	50897998	&	05 33 31.45 & -01 09 21.86	&	EB+bCep+SPB	&	5.16	&	5.34	&	1.4853(2)	&	1468.45(10)	&	9.179(3)	&	1.5(2)	&	18.9	&	B1V+B4.5V 	&	SW21	\\
VZ Cen      	&	304803692	&	11 52 28.76 & -61 31 26.93	&	EB+bCep	&	8.36	&	8.36	&	4.92870(2)	&	1574.95(20)	&	6.03025(7)	&	3.88(3)	&	67.1	&	B2III/IV       	&	IJ21	\\
\hline
\end{tabular}}
\end{center}
\tablecomments{References: (L10) \citet{Lee1910}, (J80) \citet{Jerzykiewicz1980}, (W51)  \citet{Walker1951},  (SWB2a) \citet{SouthworthandBowman2022}, (BO20) \citet{BalonaandOzuyar2020}, (DK04) \citet{Dvorak2004}, (OW05) \citet{2005IBVS.5644....1O}, (PP08) \citet{PigulskiandPojmanski2008}, (IJ21) \citet{Ijspeertetal2021}, (C22) \citet{Chenetal2022}, (LB20) \citet{Labadie-Bartzetal2020}, (F05) \citet{2005A&A...429..631F},   (BD80) \citet{1980AcA....30..501B}, (OT03) \citet{Otero2003},  (AG12) \citet{Alfonso-Garzonetal2012}, (SW20) \citet{Southworthetal2020} and (SW21) \citet{Southworthetal2021}.    SA00, BM22, DR10, SWB22b and K68 in Table 2 refer to \citet{ShibahashiandAerts2000},  \citet{Bowmanetal2022}, \citet{Drobeketal2010}, \citet{southworthandBowman2022b} and \citet{Kukarkinetal1968} respectively. These authors classified the stars either as eclipsing binary or {\ensuremath{\beta}} Cep pulsator or both. The details of their classification and other brief information about the stars are given in Appendix A.}
\end{table}

Here we present the results of the analysis of the {\it TESS} photometry of 8055 stars. We report a total of 78 binary systems out of which we classify 43 as definite eclipsing binaries containing $\beta$ Cep pulsators; a further 35 systems are listed as candidates that need further confirmation. 
There are also ten possible ellipsoidal variables with pulsating components in our sample. 
The full lists of the definite {\ensuremath{\beta}} Cep pulsators and pulsating candidates as well as ellipsoidal variables with their orbital ephemerides are shown in Tables $1-3$.  Table 1 shows the list of definite {\ensuremath{\beta}} Cep pulsators in eclipsing binaries. Stars with some unresolved blends and weak pulsations for which we were unable to unambiguously pinpoint the source of the pulsations and/or eclipses are listed as eclipsing {\ensuremath{\beta}} Cep candidates and are shown in Table 2. Table 3 lists possible ellipsoidal variables with {\ensuremath{\beta}} Cep pulsating component(s). Objects that may appear to be $\beta$~Cep pulsators in eclipsing binaries but according to our analyses are not are
shown in Table 4. Stars which were rejected at the early stage of the analysis for conspicuously showing neither eclipses nor $\beta$ Cep pulsations are not included in Table 4. The phase diagrams of light curves, and Fourier spectra of the residuals after subtracting the orbital light variations from the light curves of the stars in Tables $1-3$ are shown in Figures $4-9$ respectively. The phase diagrams of stars with incomplete ephemerides or insufficient data to give reliable phased light curves were not included in Figures 4 to 6.

The sample of pulsators also contains stars whose photometric variability had been previously identified in literature.  16 Lac \citep{Jerzykiewicz1980}, HD 101838, V4386 Sgr and V916 Cen which have been previously identified as {\ensuremath{\beta}} Cep stars in eclipsing binaries \citep[e.g.][]{PigulskiandPojmanski2008, Chenetal2022} are corroborated by {\it TESS} photometry. Several others such as HD 339003 and HD 344880 which were discovered to be $\beta$ Cep stars in eclipsing binaries in the KELT project by \citet{Labadie-Bartzetal2020},  $\lambda$ Sco from space- and ground based data \citep{2008JPhCS.118a2012B,2013A&A...557A...1H} are also corroborated by the {\it TESS} photometry. Further findings in the literature using the {\it TESS} data which are corroborated by the results of the analysis in this paper include:  V453 Cyg \citep{Southworthetal2020}, VV Ori \citep{Southworthetal2021}, V446 Cep, HD 76838 and HD 227877 \citep{Chenetal2022}, AN Dor, EO Aur, HQ CMa, QX Car, V446 Cep and V2107 Cyg \citep{SouthworthandBowman2022}, and V1388 Ori  \citet{2022Obs...142..161S}. The stars CD-38 4128, HD 303115 and Hilt 1208 have recently also been reported by \citet{Shietal2024}.  

However, some stars which were previously rejected are accepted in this paper. HD 254346, for instance, which was rejected by \citet{Labadie-Bartzetal2020} owing to blending with the $\delta$ Scuti pulsator HD 43385, 112 arc seconds away, is accepted as a {\ensuremath{\beta}} Cep star as we concluded that both HD 254346 and HD 43385 are pulsators with the former being a {\ensuremath{\beta}} Cep star. HD 329379 which was classified as an ellipsoidal variable with pulsating components by Steindl et al. (2021) is reclassified, in this paper, as a {\ensuremath{\beta}} Cep pulsator in an eclipsing binary with some ellipsoidal variations. 

On the other hand, LS CMa, HD 217919, FZ CMa and $\eta$ Ori which were classified as {\ensuremath{\beta}} Cep candidate/definite pulsators in an eclipsing binary by \cite{SouthworthandBowman2022} and  V Pup \citep{Buddingetal2021} were dropped as $\beta$ Cep pulsators in this paper. The single pulsation frequency of LS CMa is in the domain of SPB rather than  $\beta$ Cep pulsations. The pulsations of HD 217919 appear in groups of harmonically related signals with the basic frequencies in the SPB domain. Hence we do not regard the higher frequency signals to be intrinsic \citep[cf.][]{2015MNRAS.450.3015K}. FZ CMa appears to be a related case, with its pulsation spectrum dominated by a 2.7 c/d signal and possible harmonics and sub-harmonics. Also, the masses derived for the components of FZ CMa by \citet{SouthworthandBowman2022} are around 5 times the solar mass, which is too low for $\beta$ Cep stars. There are stochastic light variations present in the light curves of V Pup and possible instrumental effects (saturation), but no coherent periodic variability apart from the eclipses was found. CW Cep \citep{LeeandHong2021} is classified by us as a gravity-mode pulsator (see Appendix A). Eclipse removed light curves of $\eta$ Ori show only a single frequency near 2.31 c/d and its first harmonic. We therefore reject it as a $\beta$ Cep pulsator.

In addition, several other objects (e.g. HD 277860) that have previously been identified as eclipsing binaries \citep[e.g][]{Ijspeertetal2021} or $\beta$ Cep pulsators (e.g. Hilt 1208) \citep{Labadie-Bartzetal2020} only are now reported to show both types of variability. A brief discussion of each of the pulsators is given in Appendix A.

A number of stars which were previously classified as {\ensuremath{\beta}} Cep pulsators in eclipsing binaries, but have spectral types outside the range defined by our selection criteria or have not been observed by {\it TESS} at the time of this work were not captured in detail in this analysis. These stars include CD$-$44 4484 -- identified as a {\ensuremath{\beta}} Cep candidate in an eclipsing binary by \citet{Labadie-Bartzetal2020}, who recommended reanalysis of the target to confirm or disapprove it owing to their observation of shallow eclipses, V447 Cep - identified as {\ensuremath{\beta}} Cep by \citet{StankovandHandler2005} and eclipsing binary by \citet{Labadie-Bartzetal2020}, CD$-$46 4432 - previously identified as a {\ensuremath{\beta}} Cep pulsating candidate but rejected by \citet{Labadie-Bartzetal2020} and HD 168050 - identified as {\ensuremath{\beta}} Cep in eclipsing binary by \citet{PigulskiandPojmanski2008} but which has not been observed by {\it TESS}. At a more curious look at these targets, CD$-$44 4484 which has a spectral type B5, appears to be an eclipsing binary with {\ensuremath{\beta}} Cep pulsating component. V447 Cep turns out to be an eclipsing binary with ellipsoidal variations of different period, but no pulsations. CD$-$46 4432 which was rejected and its variability attributed to CD$-$46 4437 by \citet{Labadie-Bartzetal2020} appears to be the one pulsating instead. However, it has a spectral type of B5 and beyond the spectral range analysed in this paper.  

The number of confirmed {\ensuremath{\beta}} Cep pulsators in eclipsing binaries so far was fairly small($<30$). Using the {\it TESS} photometry, 78 definite and candidate {\ensuremath{\beta}} Cep pulsators in eclipsing binaries are reported in this work among which 43 are adjudged definite and 35 are candidate $\beta$ Cep pulsators  respectively. Nineteen of the 78 pulsators were already previously identified in literature as $\beta$ Cep pulsators in eclipsing binaries, 23 identified as eclipsing binaries only and 7 identified as $\beta$ Cep pulsators only. This results in 59 new discoveries in this work in which both variability (eclipsing binary and $\beta$ Cep pulsation) are newly identified in 29 systems and single variability (eclipsing binary or $\beta$ Cep pulsation) are newly identified in 30 stars. Ten ellipsoidal variables with $\beta$ Cep pulsating components are also identified in this work with their $\beta$ Cep variability newly identified in nine out of the ten.

 \begin{table}[ht]
\caption{List of candidates for {\ensuremath{\beta}} Cep pulsators in eclipsing binaries  with their orbital ephemerides and dominant pulsation frequencies. The dominant $\beta$ Cep pulsations and their amplitudes are given with their analytical errors. }
\begin{center}
\resizebox{\textwidth}{!}{\begin{tabular}{cccccccccccccc}
    \hline
    \multicolumn{14}{c}{}\\
        Name 	&	TIC ID	&	      RA                     & Dec      	&	Variability 	&	  Bmag  	&	  Vmag  	&	P (d)	&	T (BTJD)	&	F (d$^{-1}$)	&	A (mmag)	&	S/N	&	SpType   	&	Ref \\		
\hline																									
23 Ori       	&	264592553	&	05 22 50.00 & +03 32 40.05	&	EB+bCep?	&	4.86	&	5	&	4.56(1)	&	1470.805(20)	&	11.7727(8)	&	1.90(6)	&	43.9	&	B2IV/V         	&	IJ21	\\
$\beta$ Cep	&	321818578	&	21 28 39.59&+70 33 38.57	&	EB+bCep?	&	3.01	&	3.23	&	4.6083(9)	&	2856.233(50)	&	5.249626(4)	&	13.739(9)	&	37.3	&	B0.5IIIs 	&	SA00, BM22	\\
AN Dor 	&	220430912	&	04 52 28.24 & -55 41 49.41	&	EB+bCep/SPB?	&	7.47	&	7.69	&	2.032671(3)	&	2155.71301(1)	&	3.20275(4)	&	1.21(3)	&	27.4	&	B2/3V 	&	SWB22a	\\
CD$-$28 5247     	&	130348163	&	07 59 44.31 & -29 00 55.16	&	EB+bCep?	&	11.75	&	11.33	&	4.256(2)	&	1495.362(20)	&	4.976(2)	&	0.53(5)	&	8.9	&	B0V/Obe            	&		\\
CD$-$32 4839     	&	144535458	&	08 05 56.85 & -33 00 21.91	&	EB+bCep? 	&	12.21	&	11.57	&	4.1317(5)	&	2229.6(6)	&	5.097(1)	&	1.0(1)	&	9.8	&	B2/4           	&		\\
CD$-$32 4971     	&	145760541	&	08 12 21.08 & -32 35 03.34	&	betCep?	&	11	&	10.7	&	8.787(2)	&	2232.5(3)	&	5.3263(6)	&	1.69(9)	&	19.6	&	B1IV           	&		\\
CD$-$44 4813     	&	28699677	&	08 46 53.39 & -44 43 20.06	&	EB+bCep?	&	11.29	&	10.74	&	22.9444(5)	&	1548.92(30)	&	4.4475(7)	&	0.29(2)	&	7.7	&	B2/5           	&		\\
CD$-$57 3146     	&	463899225	&	10 19 12.87 & -58 26 01.03	&	EB+bCep?	&	10.77	&	10.73	&	23.283(8)	&	1564.925(23)	&	5.0817(3)	&	1.84(5)	&	28.9	&	B3/4           	&	IJ21	\\
GSC 04052-01378	&	390515222	&	02 53 08.35 & +62 06 10.49	&	EB+bCep?	&	     $-$  	&	11.7	&	 18.310(6)	&	1798.018(10)	&	3.0631(2)	&	5.03(8)	&	36.1	&	B2             	&		\\
HD 101838      	&	321947833	&	11 42 49.28 & -62 33 54.98	&	EB+bCep?	&	9.56	&	9.55	&	5.41187(7)	&	1574.645(10)	&	3.12772(3)	&	10.72(3)	&	19.52	&	B0.5/1III      	&	PP08, DR10	\\
HD 135477      	&	455667423	&	15 18 06.61 & -60 05 38.91	&	EB+bCep+SPB?	&	8.15	&	8.2	&	2.0079(3)	&	1626.205(60)	&	5.895(3)	&	0.22(4)	&	5.9	&	B2V            	&	IJ21	\\
HD 138112      	&	284374921	&	15 32 54.50 & -59 48 26.29	&	EB+bCep?	&	9.63	&	9.51	&	21.037(6)	&	2345.70(20)	&	5.153(1)	&	0.23(3)	&	4.97	&	B2IV-V/B5	&		IJ21\\
HD 13969       	&	264613302	&	02 17 49.85 & +57 05 25.58	&	EB+bCep?	&	9.15	&	8.0183	&	16.305(5)	&	2887.34(50)	&	7.1265(6)	&	0.62(2)	&	19	&	B0.5I          	&		\\
HD 143605      	&	427406031	&	16 04 15.99 & -56 26 27.39	&	EB+bCep?	&	9.225	&	9.148	&	2.2077(2)	&	1627.96(2)	&	4.963(4)	&	0.20(4)	&	5.2	&	B3V            	&		\\
HD 151083      	&	51288359	&	16 47 51.39 & -51 46 04.05	&	EB+bCep?	&	10.08	&	9.54	&	1.8016(4)	&	1632.945(50)	&	6.242(2)	&	1.9(2)	&	16.2	&	B2Vn           	&		\\
HD 154407      	&	42424196	&	17 06 47.90 & -35 53 15.07	&	EB+bCep?	&	9.12	&	8.83	&	2.5592(4)	&	1631.54(10)	&	12.968(2)	&	0.35(4)	&	10.6	&	B2Vn           	&		\\
HD 155274      	&	45731009	&	17 11 57.77 & -35 00 26.32	&	EB+bCep+SPB?	&	10.14	&	9.73	&	6.896(7)	&	1629.9744(50)	&	10.143(1)	&	1.15(5)	&	37.6	&	B2/5(n)        	&		\\
HD 1743        	&	428415975	&	00 22 03.41 & +62 11 06.29	&	EB+bCep?	&	8.43	&	8.33	&	11.7950(5)	&	1801.35(36)	&	8.7933(6)	&	0.24(6)	&	12.9	&	B0.2IV         	&	IJ21	\\
HD 240171      	&	314034024	&	23 02 44.51 & +57 08 33.51	&	EB+bCep?	&	10.02	&	9.92	&	5.1514(9)	&	1630.75(10)	&	7.65284(7)	&	1.29(4)	&	37.8	&	B1V            	&		\\
HD 277680      	&	122883754	&	05 10 17.68 & +40 39 35.26	&	EB+bCep?	&	8.93	&	8.89	&	4.3207(5)	&	1818.8(2)	&	8.4876(9)	&	0.95(4)	&	28	&	B3             	&	IJ21	\\
HD 306124      	&	466887289	&	11 08 51.19 & -61 21 10.48	&	betCep?	&	10.89	&	10.68	&	1.4074(8)	&	2310(4)	&	5.545(5)	&	0.12(3)	&	4.9	&	B3             	&		\\
HD 46060       	&	25041731	&	06 30 49.81 & -09 39 14.80	&	EB+bCep?	&	9.03	&	8.82	&	20.6094(8)	&	1474.22(20)	&	9.9603(9)	&	0.38(1)	&	25.6 &	B2II           	&		\\
HD 73903       	&	141497319	&	08 38 58.88 & -46 13 37.55	&	EB+bCep?	&	9.19	&	9.05	&	2.8093(2)	&	1520.19(20)	&	4.25210(7)	&	15.19(9)	&	94.9	&	B3II           	&	\\
HD 76838 	&	30562668	&	08 57 07.55 & -43 15 22.27	&	EB+bCep?	&	7.31	&	7.31	&	3.8522(2)	&	1518.08(40)	&	8.7329(1)	&	2.84(3)	&	93.9	&	B2IV 	&	C22	\\
HD 92741       	&	458561474	&	10 41 12.35 & -59 58 25.04	&	EB+bCep?	&	7.22	&	7.25	&	5.373(2)	&	1574.57(20)	&	4.9674(1)	&	2.72(4)	&	47.7	&	B1/2II         	&		IJ21\\
HD 92782       	&	458599043	&	10 41 30.95 & -57 30 51.61	&	EB+bCep?	&	9.62	&	9.51	&	14.46(2)	&	2287.375(20)	&	6.52581(6)	&	12.02(3)	&	65.9	&	B1/2(III)      	&		\\
Hilt 1208	&	269228628	&	23 17 13.69&+60 00 27.93	&	EB+bCep?	&	11.35	&	10.9	&	2.422(1)	&	2856.38(4.00)	&	4.99209(4)	&	8.04(3)	&	120.4	&	B2 	&	LB20	\\
LS VI -04 19   	&	33091633	&	06 59 30.23 & -04 48 43.82	&	EB+bCep?	&	11.1	&	10.81	&	7.520(4)	&	1495.83(30)	&	10.564(1)	&	0.51(2)	&	20.5	&	B0IV           	&		\\
NGC 2483 2     	&	129364361	&	07 55 43.58 & -27 54 05.42	&	EB+bCep+SPB?	&	12.46	&	12.16	&	4.767(5)	&	1493(3)		& 4.162(1)	& 	1.9(2)		& 11.56		&B3    	\\
QX Car 	&	469247903	&	09 54 33.88 & -58 25 16.58	&	EB+bCep?	&	6.47	&	6.64	&	4.47800(2)	&	1546.32(10)	&	5.722(2)	&	0.13(2)	&	5.5	&	B2V 	&	SWB22a	\\
TYC 8995-5052-1	&	314139276	&	13 30 00.85 & -63 38 04.24	&	EB+bCep?	&	11.77	&	11.43	&	4.73(1)	&	1665.5301(40)	&	5.896(3)	&	12.4(1.7)	&	16.9	&	B2III 	&		\\
V1388 Ori   	&	337165095	&	06 10 59.17 & +11 59 41.48	&	EB+bCep+SPB?	&	7.44	&	7.5	&	2.1870301(8)	&	2485.226609(16)	&	3.9987(8)	&	0.18(5)	&	5	&	B2V            	&	SWB22b	\\
V446 Cep    	&	335265326	&	22 08 45.59 & +61 01 20.71	&	EB+bCep?	&	7.4	&	7.32	&	3.808386(6)	&	1767.0005(1)	&	10.24378(1)	&	1.280(7)	&	93.5	&	B1V            	&	IJ21,SWB22a	\\
V536 Mon    	&	23548300	&	07 13 55.57 & -02 54 29.87	&	EB+bCep+SPB?	&	9.29	&	9.41	&	6.1341(3)	&	1493.21(20)	&	6.541(2)	&	0.44(3)	&	20.5	&	B3/4III        	&	K68	\\
V964 Sco    	&	339567760	&	16 54 18.32 & -41 51 35.65	&	EB+bCep?	&	9.776	&	9.603	&	5.2179(5)	&	2366.63(20)	&	11.3812(3)	&	1.70(3)	&	23.4	&	B0.5V          	&		\\

\hline
\end{tabular}}
\end{center}
\end{table}

\subsection{Periodicities of the pulsators\label{subsec:figures}}

 The orbital periods of the stars and their analytical uncertainties are given in Tables $1-4$. Orbital periods available in literature or obtained with PHOEBE Estimator \citep{Prsaetal2016} are used as inputs for the fitting of the harmonic model and estimation of precise  orbital period  using Period04 \citep{LenzandBreger2005} during prewhitening. Where there are already established precise periods in the literature, we adopted them after checking their validity for the times of {\it TESS} observation. The validity of the orbital period values was checked by comparison with a multiharmonic fit to the orbital light curves optimized within {\it Period04}. 
 For stars with incomplete orbital cycles in the 30-minute light curve (i.e. light curves with only the primary eclipse visible and with gaps between the consecutive primary eclipses), we determined the periods with the 2-minute cadence light curves if available. In the QLP data, sections of data that are present in the 2-minute SPOC data are sometimes missing.  Where there is no light curve with a complete orbital cycle in both cadences (2-minute SPOC and QLP data) a lower limit to the period is adopted.
 The analytical uncertainties of the periods were estimated using the {\it Period04} program.  We obtained an orbital period range of 1.4074--161.25 d for our sample of pulsators of 78 definite and candidate $\beta$ Cep pulsators. However, more than 80$\%$ of the sample has orbital periods below 20 d.

Period changes and modulations are usually examined via eclipse timings and O-C diagrams. However, the time span of the {\it TESS} data, which is in most cases only one or two months, makes it unreliable to compute them  as they are too short to show any period change. As a result, the eclipse timings and O-C diagrams are not computed. Nevertheless, we computed the epochs of the binary light curve by estimating the time of minimum (i.e. the midpoint) of the primary eclipse, which in binary star analysis can also be referred to as the time of superior conjunction (t$\_$supconj). We arrived at the epochs by manually zooming into the primary eclipses and reading off the midpoint of the eclipse curve. The errors were estimated visually on a case-by-case basis and are thus upper limits to the real error.

\begin{table}[ht]
\caption{List of ellipsoidal variables with $\beta$ Cep pulsating components.  The dominant $\beta$ Cep pulsations and their amplitudes are given with their analytical errors. }
\begin{center}
\resizebox{\textwidth}{!}{\begin{tabular}{ccccccccccccc}
    \hline
    \multicolumn{13}{c}{}\\
        Name	&	TIC ID	&	      RA                     & Dec      	&	Variability 	&	  Bmag  	&	  Vmag  	&	P (d)	&	F (d$^{-1}$)	&	A (mmag)	&	S/N	&	SpType   	&	Ref\\	
     \hline																							
BD+48 658      	&	301263808	&	02 23 23.60 & +49 01 55.41	&	EB/ELL+BCep	&	8.65	&	8.76	&	2.955(2)	&	6.1044(8)	&	1.74(6)	&	37.7	&	B2             	&		\\
HD 13338       	&	347486043	&	02 12 19.17 & +57 56 27.17	&	ELL/ROT+bCep	&	9.38	&	9.17	&	 $-$	&	4.63533(9)	&	5.548(9)	&	11.8	 &	O9.5V/B1III         	&		LB20\\
HD 232874      	&	266338052	&	04 02 15.74 & +53 45 11.78	&	EB/ELL+bCep	&	9.26	&	8.92	&	1.5174(8)	&	5.7395(3)	&	7.59(9)&	11.2	&	B0.5V          	&	LB20	\\
HD 277132      	&	121064859	&	04 56 27.87 & +41 16 18.94	&	EB/ELL/ROT+bCep	&	11.58	&	11.23	&	2.4291(4)	&	4.621(4)	&	0.7(1)	&	9.3	&	B3             	&		\\
HD 300978      	&	458604640	&	10 41 55.77 & -56 42 51.88	&	EB/ELL/ROT+bCep	&	9.66	&	9.61	&	4.787(8)	&	7.6876(1)	&	1.69(2)	&	31.2	&	B3             	&		\\
HD 308106      	&	465870314	&	10 58 11.89 & -62 09 36.41            	&	ROT/ELL/EB+bCep	&	10.77	&	10.65	&	1.8837(6)	&	13.709(1)	&	0.36(5)	&	6.6	&	B3             	&		\\
HD 327010      	&	382486122	&	17 11 42.98 & -42 52 02.75	&	bCep+ELL? 	&	10.36	&	9.8	&	7.5537(6)	&	4.7824(9)	&	2.48(9)	&	27.8	&	B3             	&		\\
HD 39716       	&	66975228	&	05 54 03.33 & -06 45 08.27	&	ELL?+SPB?+bCep?	&	8.5	&	8.52	&	2.4903(3)	&	5.4376(9)	&	0.39(2)	&	24.9	&	B3III          	&		\\
HD 55687       	&	178374964	&	07 13 34.03 & -10 29 25.34	&	EB/ELL+SPB?+bCep	&	9.27	&	9.34	&	1.211(1)	&	7.82(1)	&	0.4(5)	&	7.6	&	B3II/III       	&		\\
NGC 6913 82    	&	14621767	&	20 24 50.37  & +38 18 34.60	&	EB+bCep	&	12.96	&	11.53	&	2.0499(1)	&	6.9686(2)	&	2.46(4)	&	34.9	&	B2IV            	&		\\

\hline
\end{tabular}}
\end{center}
\end{table}

 \begin{table}[ht]
\caption{List of rejected targets.}
\begin{center}
\resizebox{\textwidth}{!}{\begin{tabular}{cccccccccc}
    \hline
    \multicolumn{8}{c}{}\\
       Name 	&	TIC ID	&	      RA                     & Dec      	&	  Bmag  	&	  Vmag  	&	P (d)	&	SpType  \\
       \hline
$\eta$ Ori 	&	4254645 	&	05 24 28.62 & -02 23 49.73		&	3.18 	&	3.35 	&	7.989(3)	&	B1V+B2 	\\       
BD+61 675      	&	84342607                       	&	04 07 44.08 & +62 18 04.13	&	10.16	&	9.61	&	2.699(7)	&	B1Vn           	\\
CD$-$59 4169     	&	383089500                      	&	12 14 33.31 & -60 24 40.92	&		11.15	&	10.72	&	2.97154(9)	&	B1V\\
CPD$-$59 3141    	&	467066902                      	&	11 11 14.84 & -60 29 51.08	&		11.25	&	11.06	&	11.5089(4)	&	B2      	\\
CW Cep 	&	 434893323 	&	23 04 02.23 & +63 23 48.72	&		7.97 	&	7.6 	&	2.72914(3)	&	B1.5Vn 	\\

FZ CMa 	&	 125497512 	&	07 02 42.61 & -11 27 11.57	&		8.28 	&	8.14 	&	1.2733(3)	&	B2IVn 	\\

HD 153772      	&	212415990                      	&	17 04 01.23 & -51 05 01.15	&	8.38	&	8.32	&	-	& B2V  \\
HD 154646      	&	380870688                      	&	17 09 01.58 & -46 16 38.61	&	9.96	&	9.64	&	13.0298(1)	&	B2II           	&		\\
HD 217919 	&	 434723918	&	23 03 01.47 & +63 41 53.35	&		8.75 	&	8.2 	&	16.2080(9)	&		B3 	&	\\
HD 308111      	&	465869053                      	&	10 57 54.75 & -62 17 35.73	&	10.62	&	10.59	&	2.4396(5)	&	B2             	\\
LS II +23 34   	&	360661624                      	&	19 44 21.08 & +23 17 05.90	&	12.3	&	11.82	&	10.2572(2)	&	B              	\\
LS CMa & 63427664 & 07 01 05.95 & -25 12 56.28 &  5.47 &  5.64 & $>$ 10 & B2/3III/IV \\

V Pup 	&	 269562415 	&	07 58 14.44 & -49 14 41.68	&		4.24 	&	4.41 	&	1.45453(3)	&	B1Vp+B2 	\\

\hline
\end{tabular}}
\end{center}
\end{table}


\section{Conclusion} \label{sec:cite}
In this paper, we conducted a search for {\ensuremath{\beta}} Cep pulsators in eclipsing binaries in a sample of 8055 stars of spectral type B0--B3 using the {\it TESS} QLP and 2-minute cadence data. The result of the photometric and pulsation analyses indicates a total of 78 pulsators in eclipsing binaries in which 43 are recognized as definite and 35 are candidate pulsators after accounting for blends and removal of false positives. We computed the orbital ephemerides and tabulated the dominant pulsation frequencies with their amplitudes and S/N. We further crossmatched our results with the results of previous searches. Among the 78, 59 are new discoveries as 19 have been previously identified as {\ensuremath{\beta}}  Cep pulsators in eclipsing binaries. 23 stars  have also been previously identified as eclipsing binaries only and 7 as {\ensuremath{\beta}} Cep pulsators only. The number of pulsators in our sample accounts for about 1\% of the sample and contributes about 59 new definite {\ensuremath{\beta}} Cep pulsators and candidates in eclipsing binaries to the pool already discovered to date. There are also 10 possible ellipsoidal variables with pulsating components in our sample.  This work provides a bigger sample for a more general and homogeneous in-depth asteroseismic analysis of {\ensuremath{\beta}} Cep pulsators. It will provide the needed constraints to better calibrate the internal mixing profile and possibly resolve the mass discrepancy in massive stars via the combined strengths of binary and asteroseismic modelling.

%
%
%
%
\section{Acknowledgments}
\begin{acknowledgments}
This work was supported by the Polish National Science Foundation (NCN) under grant nr. 2021/43/B/ST9/02972.
This paper used the {\it TESS} data obtained from the Mikulski Archive for Space Telescopes (MAST). Support to MAST for these data is provided by the NASA Office
of Space Science via grant NAG5-7584 and by other grants and contracts. Funding for the {\it TESS} mission is provided by the NASA Explorer Program. This paper also made use of the SIMBAD database and the VizieR catalogue access tool, operated at CDS, Strasbourg, France; and the SAO/NASA Astrophysics Data System. It also made use of data from the European Space Agency (ESA) mission
{\it Gaia} (\url{https://www.cosmos.esa.int/gaia}), processed by the {\it Gaia}
Data Processing and Analysis Consortium (DPAC,
\url{https://www.cosmos.esa.int/web/gaia/dpac/consortium}). Funding for the DPAC
has been provided by national institutions, in particular the institutions
participating in the {\it Gaia} Multilateral Agreement. We also thank the referee for their invaluable constructive review and recommendations that helped to improve the paper.
\end{acknowledgments}

%

\vspace{5mm}
\facilities{{\it TESS}}

The data presented in this paper were obtained from the Mikulski Archive for Space Telescopes (MAST) at the Space Telescope Science Institute.  The observations analysed for the stars in Tables1-4 in the paper can be accessed via \dataset[DOI]{https://doi.org/10.17909/8rpk-re37} 

\software{python,  
          Period04 \citep{LenzandBreger2005}, 
          Eleanor \citep{Feinsteinetal2019,Brasseuretal2019, Burkeetal2020}}, {\it TESS}-Localize \citep{HigginsandBell2022}



\appendix 

\section{Short remarks on individual stars}

\subsection{{\ensuremath{\beta}} Cep pulsators in eclipsing binaries\label{subsec:figures}}

16 Lac       	=	TIC 129538133: It was discovered to be a spectroscopic binary by \citet{Lee1910}, a pulsating star by \citet{Walker1951} and an eclipsing binary by \citet{Jerzykiewicz1980}. Detailed studies of the star to determine its orbital and absolute parameters have been undertaken by a number of authors \citep[e.g.][]{2001A&A...367..236L,SouthworthandBowman2022}.  An asteroseismic analysis of the pulsating component was conducted by \cite{2003A&A...406..287T}.\\

$\lambda$ Sco 	=  TIC 465088681 is a multiple system with an inner eclipsing binary and a tertiary component. The inner eclipsing binary is composed of a B-star of mass $10.4\pm1.3$ solar mass showing $\beta$ Cep pulsations and a $1.6-2.0$ solar mass unresolved main sequence star \citep{SouthworthandBowman2022}. Detailed analyses of the system can be found over a wide range of literature \citep[e.g][]{1997A&A...324.1096D, 2013A&A...557A...1H}.  \\

BD+44 3594     	=	TIC 353099086: this is a Be star \citep{JaschekandEgret1982} with a long orbital period. Although it has a long period variable candidate (2MASS J20490628+4525007) 59.79 arc seconds away, there is no indication of a significant blending effect. It shows both $\beta$ Cep and SPB pulsations. The binarity and the pulsations are reported for the first time in this work. \\

BD+44 3664     	=	TIC 330081196: This star has its spectral type classified as B1Vn by \citet{Straizysetal1989}. It is an eclipsing binary with $\beta$ Cep pulsations and has some nearby Gaia sources which could contaminate the light curve. However, the CROWDSAP parameter (i.e. the crowding metric that reflects the fraction of the flux in the aperture that is due to the target itself not the nearby sources.) for this object is 0.89624, suggesting that the signals most likely originate from the target. The binarity and the pulsations are newly identified in this work. 
\\ 

CD$-$38 4128     	=	TIC 134522557: This star has no significant nearby contaminator. It has been observed in {\it TESS} sectors 6, 7, 34,35, 61 and 62 using {\it TESS} camera 3. Whereas the binarity was first reported by \citet{BalonaandOzuyar2020}, its $\beta$ Cep variability is newly reported in this work. There are still a number of pulsations left in its phase diagram shown in Figure 3.\\

CD$-$51 9984     	=	TIC 314529804: Classified as a B0.5III star in the spectral classification and photometry of southern B stars compiled by \citet{Feastetal1961}. This star has a neighbour, TIC 314529900, which is 1 magnitude brighter at 76.18 arc seconds from it and claimed to be an eclipsing binary star \citep{Ijspeertetal2021}. The available {\it TESS} photometry of the latter star does however not show any eclipses. Therefore, we suggest that the eclipses and pulsational signals we observed emanate from CD$-$51 9984. However, its $\beta$ Cep variability was first identified by \citet{BalonaandOzuyar2020}.    \\   

CPD$-$41 7746    	=	TIC 339570153: There are many faint nearby sources within the aperture mask (e.g. 2MASS J16542911-4139115). However, the CROWDSAP parameter for this object is 0.89506, suggesting that the variability signals most likely originate from the target. \\ 
                   
CPD$-$45 3109    	=	TIC 28957011:  This star has no nearby contaminant. The binarity and the pulsations are reported for the first time in this work.  \\

CZ Vel      	=	TIC 355656323 was classified as an eclipsing binary by \citet{Dvorak2004}. Pulsations are reported here for the first time.   \\ 

EK Cru      	=	TIC 379012185 was identified as an eclipsing binary by \citet{2005IBVS.5644....1O}; pulsations (both of the SPB and $\beta$ Cephei type) are reported here for the first time.  \\

EO Aur 	=  TIC 408937625: It was  discovered to be eclipsing by \citet{Gaposchkin1943} and to be $\beta$ Cep pulsator by \citet{SouthworthandBowman2022}. It contains two early-B stars \citep{SouthworthandBowman2022} in a 4.06550(3) d orbit.\\

HD 108628      	=	TIC 450918869: It was discovered to be a {\ensuremath{\beta}} Cep pulsator by \citet{PigulskiandPojmanski2008}. These authors, however, treated it as an isolated single star as no eclipses seem to have been detected by them. However the {\it TESS} QLP light curve shows primary eclipses of 18 mmag depth. \\   

HD 112026      	=	TIC 436382800: Although there is a nearby star TIC 436373277 = HD 312121, which could possibly modulate the light curve,  the pulsational signals we observed are traced to HD 112026. The CROWDSAP parameter for the target is  also 0.96683, and  the amplitude of the dominant pulsation is  2.22 mmag. The orbital period of some 43.205(9) d is rather long and the orbit is clearly eccentric. The binarity and the pulsations are reported for the first time in this work.\\

HD 112485      	=	TIC 437617380: It was identified as an eclipsing binary in the catalogue of OBA-type eclipsing binaries observed by {\it TESS} compiled by \citet{Ijspeertetal2021}. Our analysis suggests that there is no significant contamination from TIC 437612835, which is an A0 star 89.69 arc seconds away. \\  
            
HD 113742      	=	TIC 440817830:  This star has a relatively long orbital period of 15.68844(8) d and an eccentric orbit. There is a nearby G3V star, with a similar {\it TESS} magnitude, at 106.85 arc seconds from it. However, our blending analysis suggests that the detected light variations come from HD 113742. The eclipses and pulsations are reported for the first time here.  \\

HD 150927      	=	TIC 78636551: It was discovered to be a {\ensuremath{\beta}} Cep variable by \citet{PigulskiandPojmanski2008}. The {\it TESS} light curves imply the presence of eclipses with depths of about 0.04 mag and a 14.33(2) d period within the high-amplitude pulsations (the orbital period could be that or twice that value). The dominant pulsation frequency of 4.84325(2) d$^{-1}$ appears to be within an unresolved multiplet, hence our determination of its amplitude is uncertain (but in relatively good agreement with \citet{PigulskiandPojmanski2008}). \\

HD 151791      	=	TIC 246552414: Two stars (SAO 227312 and TYC 7880-2706-1), which are about 2.5 magnitude fainter than the target are found within 36 arc seconds from the target. Our blending analysis suggests that the signals are consistent with location of the brightest star, but not uniquely so. The star 36 arc seconds  away also seems to be a hot pulsating star. The eclipses and pulsations are reported for the first time here.    \\  

HD 152268      	=	TIC 339680203: also identified as eclipsing binary in the catalogue of OBA-type eclipsing binaries observed by {\it TESS} compiled by \citet{Ijspeertetal2021}. Its {\ensuremath{\beta}} Cep variability is, however, newly identified in this work.    \\

HD 157400      	=	TIC 158688754: This star has no nearby contaminant. The binarity and the pulsations are reported for the first time in this work.    \\   

HD 188891      	=	TIC 171502734:  The catalogue of {\it Kepler} eclipsing binaries  \citep{Kirketal2016} lists this system with a period of 161.25 d. Pulsations are present at very low amplitude in both the SPB and $\beta$ Cephei frequency regions, both in the {\it Kepler} and {\it TESS} data.\\

HD 227877  =  TIC 91111448: It was discovered to be eclipsing by \citet{Ijspeertetal2021} and to be a $\beta$ Cep pulsator by \citet{Chenetal2022}. However, its orbital period was not computed by these authors. We obtained a period of 1.70664(4) d for the system. \\

HD 254346      	=	TIC 426520557  was rejected by \citet{Labadie-Bartzetal2020} as a $\beta$ Cep pulsator owing to blending with the $\delta$ Scuti pulsator HD 43385, 112 arc seconds away from it. However, it is accepted as {\ensuremath{\beta}} Cep in this paper as the {\it TESS} photometry allows us to conclude that both HD 254346 and HD 43385 are pulsators with the former being a {\ensuremath{\beta}} Cep star.  \\  

HD 303115      	=	TIC 458263480: there are many faint sources in the field, 7 mag fainter than the target in the G band and with the closest at 70.19 arc seconds. Our analysis of possible contamination suggests that both the eclipses and the pulsations come from the target. \\

HD 329379      	=	TIC 122314621: \citet{Steindletal2021} classified it as an ellipsoidal variable that also shows pulsations. We suggest those to be of the {\ensuremath{\beta}} Cephei type. The pulsation amplitudes are enhanced near the ellipsoidal light minima, suggesting interplay between the pulsations and tidal forces. A phase diagram of the pulsation-removed orbital light curve shows the presence of shallow eclipses.  \\  

HD 339003      	=	TIC 10891640: It was discovered to be a {\ensuremath{\beta}} Cep pulsator in an eclipsing binary by \citet{Labadie-Bartzetal2020}, who also noticed a reflection effect in the light curve. This is corroborated by the {\it TESS} photometry.\\    

HD 344880      	=	TIC 451932686:  also this star was discovered to be a {\ensuremath{\beta}} Cep pulsator in an eclipsing binary by \citet{Labadie-Bartzetal2020} which is confirmed by our analysis of its {\it TESS} photometry. It has a long orbital period of 54.49399(1) d. \\

HD 92024       	=	TIC 458076434: It is a known $\beta$ Cephei pulsator in an eclipsing binary, e.g., see \cite{2005A&A...429..631F}.\\

HD 96355       	=	TIC 466528132:  Also discovered to be an eclipsing binary by \citet{Ijspeertetal2021} in their catalogue of OBA-type eclipsing binaries observed by {\it TESS}. A source, TYC 8958-3680-1, which is 4.1 mag fainter than the target in the {\it TESS} band is at 38.58 arc seconds from it. However, the pulsational signals appear to originate from HD 96355.     \\   

HQ CMa 	=	 TIC 106830354: It was discovered to be eclipsing by \citet{JerzykiewiczandSterken1977} and as a $\beta$ Cep and SPB candidate by \citet{SouthworthandBowman2022}. The orbital period is unknown as there is no agreement in the literature about it. \citet{Sterkenetal1985} reported an orbital period of 24.6033 d whereas \citet{SouthworthandBowman2022} reported a period range of 21.2- 22.6 d or 34.5 d or longer. In this work, we also could not unambiguously determine this period. Pulsations are found both in the SPB and $\beta$ Cep frequency domains. \\

LR Ara      	=	TIC 447530589: This star was catalogued as an eclipsing binary by \citet{1980AcA....30..501B}. Although there is a contaminator, UCAC4 143-192850,  31.26 arc seconds away, which is about 3 magnitude fainter, there is no indication that the signals come from the contaminator: the eclipses are too deep that they could be caused by blending with this star which is also too cool to pulsate with the observed periods. Pulsations (both of the SPB and $\beta$ Cephei type) are reported here for the first time. \\

LS I +61 145   	=	TIC 406965391:  Also identified as an eclipsing binary in the catalogue of OBA-type eclipsing binaries observed by {\it TESS} compiled by \citet{Ijspeertetal2021}. Its {\ensuremath{\beta}} Cep variability is, however, newly identified in this work. \\ 

LS I +63 36    	=	TIC 359042331: A faint star TIC 359042198  with a magnitude difference of 4.86 in the G band is 109.94 arc seconds away from the target. The {\it TESS} light curves of TIC 359042331 show a primary eclipse depth of 34.5 mmag, a secondary eclipse depth of 18 mmag and an amplitude of the dominant pulsation of 1.05 mmag. Considering the magnitude difference between the stars, it is unlikely that the contaminator caused signals of such amplitude. We therefore conclude that both the eclipses and pulsations originate from TIC 359042331. \\ 

TYC 3699-160-1 	=	TIC 245470639: Both $\beta$ Cephei and SPB pulsations as well as a reflection effect are present. A star (TYC 3699-874-1) 0.3 mag fainter than the target in the {\it TESS} band at 69.89 arc seconds distance is present, but both the $\beta$ Cep pulsations and the eclipses arise from the target.   \\

TYC 4050-2830-1	=	TIC 458879750:   There is no nearby contaminator bright enough to significantly modulate the light curve. Both $\beta$ Cephei and SPB pulsations are present.   \\ 

V1061 Cen   	=	TIC 334443373: \citet{Otero2003} classified it as eclipsing binary. Pulsations are reported here for the first time. \\

V1166 Cen   	=	TIC 443262289 was first classified as an eclipsing binary by \citet{Alfonso-Garzonetal2012} and corroborated by \citet{Ijspeertetal2021} in their catalogue of OBA-type eclipsing binaries observed by {\it TESS}. Pulsations are reported here for the first time.    \\ 

V1216 Sco   	=	TIC 247315421:  It was identified as an eclipsing binary by \citet{Otero2003}. Pulsations are reported here for the first time.  \\

V2107 Cyg 	=	 TIC 42244951: Also discovered to be eclipsing binary  by \citet{Kazarovetsetal1999} using {\it Hipparcos} satellite data and as a $\beta$ Cep pulsator by \citet{SouthworthandBowman2022}. Rotational splitting and tidally perturbed modes are also suspected by the latter authors.  \\

V4386 Sgr      	=	TIC 60433558:  It was first discovered to be eclipsing by \citet{PigulskiandPojmanski2008}, corroborated by \citet{Avvakumovaetal2013} and also as a {\ensuremath{\beta}} Cep pulsator by \citet{PigulskiandPojmanski2008} and corroborated by \citet{Chenetal2022}.\\

V453 Cyg 	= TIC 90349611:  This eclipsing system has an orbital period of 3.8900(4) d, an eccentric  orbit and exhibits apsidal motion \citep{1973A&A....25..157W}. Its $\beta$ Cep pulsations were analysed by \citet{Southworthetal2020}, who also reported tidally perturbed pulsations in the system.   \\

V916 Cen	=	TIC 322078735: This star was discovered to be a $\beta$ Cep pulsator in an eclipsing binary with an orbital period of 1.4632361(21) d by \citet{PigulskiandPojmanski2008}. The {\it TESS} photometry in this analysis confirms its variability type.  It is also a Be star \citep{MoffatandVogt1975}, a member of the open cluster Stock 14  and shows other types of variability related to $\gamma$ Cas and $\lambda$ Eridani \citep{PigulskiandPojmanski2008}. Although there could be significant third light contribution, the amplitude of the dominant  $\beta$ Cep pulsation is reasonably large (13.2 mmag) that we cannot conclude that it originates from the modulation of light curves by the contaminating light.\\

VV Ori 	=	 TIC 50897998: It was discovered to be an eclipsing binary with $\beta$ Cep primary and an SPB secondary by \citet{Southworthetal2021}. These authors also observed a changing orbital inclination which was adjudged to be driven by its dynamical interactions with a third body. This system was also among the five $\beta$ Cep pulsators identified by \citet{Chenetal2022}. \\

VZ Cen      	=	TIC 304803692:  It was first identified as an eclipsing binary by \citet{Alfonso-Garzonetal2012}. Pulsations are reported here for the first time.    \\

\subsection{Candidates for {\ensuremath{\beta}} Cep pulsators in eclipsing binaries\label{subsec:figures}}

23 Ori       	=	TIC 264592553: This star was first identified as a spectroscopic binary by \citet{EggletonandTokovinin2008} and as an eclipsing binary by  \citet{Ijspeertetal2021}. In this work, we newly identify {\ensuremath{\beta}} Cep variability in this system. However, whereas the pulsations originate from 23 Ori, it is not clear whether the eclipses seen in the light curve come from it or from HD 35148, 32 arc seconds apart from the target. \\ 
 
$\beta$ Cep	=	TIC 321818578: It has been studied extensively as a single star \citep[][]{ShibahashiandAerts2000, Bowmanetal2022} as none of the authors had observed eclipses in its light curve. However, the {\it TESS} photometry shows shallow eclipses of 10 mmag in the QLP data and 20 mmag in sector 58 data, for instance. It is not certain that the eclipses involve the pulsating star. It could possibly come from $\beta$ Cep B which is 13.38 arc seconds from it with a V magnitude of 8.63.  \\

AN Dor =  TIC 220430912:  AN Dor was first discovered to be an eclipsing binary by \citet{Kazarovetsetal1999} using the {\it Hipparcos} satellite and catalogued as a variable B-star in eclipsing binary by \citet{PercyandAu-Yong2000}, who pointed out that the short term variability is uncertain. \citet{SouthworthandBowman2022} using {\it TESS} data reported a total eclipse and a strong reflection effect, and also observed SPB and {\ensuremath{\beta}} Cep pulsations over a wide frequency range of 1--4 d$^{-1}$ for dominant pulsations.  We concur with this finding with the caveat that we cannot confidently rule out that the variations in the $\beta$ Cep frequency domain are due to combination frequencies of the SPB pulsations.  \\

CD$-$28 5247     	=	TIC 130348163: It has weak pulsations and blends (amongst others) with TIC 130348237 which could be the origin of the pulsational signals.\\

CD$-$32 4839     	=	TIC 144535458: We detected ellipsoidal light variations and shallow eclipses. There is a single possible $\beta$ Cep pulsation mode, but given the presence of nearby sources (e.g., Gaia DR2 554761636497853056, 3.858 mag fainter than the target in the G-band, 50 arc seconds away and TYC 7125 2876-1, 0.03 mag fainter than the target in B band, 112.32 arc seconds away), we cannot be sure that they arise in the photometric binary. \\

CD$-$32 4971  = TIC 145760541: The star is an optical triple system and located in a crowded field. We cannot safely conclude that both the pulsational signals and the eclipses arise from one and the same object.\\

CD$-$44 4813     	=	TIC 28699677: This star has a long orbital period with weak pulsations that we cannot safely exclude to originate from some fainter star nearby. \\

CD$-$57 3146     	=	TIC 463899225: There are some blending issues with a few contaminators nearby. This object was discovered to be an eclipsing binary by \citet{Ijspeertetal2021} in their catalogue of OBA-type eclipsing binaries observed by {\it TESS}. The pulsations which are suspected to originate from the target are reported in this work for the first time.   \\

GSC 04052-01378	=	TIC 390515222:  We cannot exclude confidently that the weak pulsational signal detected arises in one of the neighbouring stars.\\

HD 101838      	=	TIC 321947833: Its variability was discovered by \citet{PigulskiandPojmanski2008} and {\ensuremath{\beta}} Cep pulsations by \citet{Drobeketal2010}. It was shown that the orbital period could be about 5.41167 d or twice as much \citep{Drobeketal2010}. These authors however favoured twice of 5.41167 d which could result in double eclipse of equal depths instead of single eclipse with 5.41167 d. In this paper, an orbital period of 5.41187(7) d was obtained from the {\it TESS} light curve. from the S10, 11, 37 and 38 data.  Also, the pulsation frequencies of the star are all below 3.2\,\cd, making us suspect this is in fact a rapidly rotating SPB and not a $\beta$ Cep star.\\

HD 135477      	=	TIC 455667423: It was also identified as an eclipsing binary in the catalogue of OBA-type eclipsing binaries observed by {\it TESS} compiled by \citet{Ijspeertetal2021}. The orbit is clearly eccentric. It has a faint long period variable, about 4 magnitude dimmer, at 29 arc seconds from it. Our analysis suggests that the target has low amplitude pulsations both in the SPB and $\beta$ Cephei frequency domains. As some (but not all) of the frequencies in the $\beta$ Cephei range can be explained with combinations of the lower frequencies, we retain this system among the candidates instead of firm detections. \\

HD 138112      	=	TIC 284374921:  It was identified as an eclipsing binary in the catalogue of OBA-type eclipsing binaries observed by {\it TESS} compiled by \citet{Ijspeertetal2021}. It has few contaminators nearby. The four brightest contaminators in the {\it TESS} band are three red giants and one A-type star. The eclipses are shallow and the pulsations are weak although in the $\beta$ Cep domain.  \\

HD 13969       	=	TIC 264613302:  This star has a long orbital period of about 16 d and shallow  flat bottomed eclipses of about 0.02 mag depth. The CROWDSAP parameter is 0.90977 and the amplitudes of the pulsations are below 0.6 mmag. We thus cannot be sure that both the pulsations and eclipses come from the target. \\

HD 143605      	=	TIC 427406031: Although there is a long period variable (2MASS J16041100-5625180) 80.72 arc seconds away (3.4 mag fainter than the target in the {\it TESS} passband), our blending analysis suggests that the signals emanate from HD 143605. The pulsation is of low amplitude, but significantly present in the data. However its position in the theoretical HR-diagram with a mass of $5M_\odot$ casts makes us doubt whether it is $\beta$ Cep star. We could face SPB pulsation shifted by rotation into the $\beta$ Cep regime.  \\  

HD 151083      	=	TIC 51288359 was classified as a Be star by \citet{BalonaandOzuyar2021}. The shape of light curves resembles eclipses with a short period (1.8016(4) d or half its value) with pulsations superposed. The features reminiscent eclipses cannot be fully distinguished from the beating of multiple pulsation modes. \\ 

HD 154407      	=	TIC 42424196 is a visual double with TIC 42424192 that has almost identical basic stellar parameters. The pulsations and eclipses could thus arise from different objects, respectively, or from either of the two stars.\\

HD 155274      	=	TIC 45731009: It is 5.71 arc seconds away from another eclipsing binary, TIC 45731014 first identified by \citet{Ijspeertetal2021}.  It is not clear where the variability observed in both is coming from: either both are eclipsing binaries or one is modulated by the other. We cannot rule out that \citet{Ijspeertetal2021} misidentified the source of the eclipses. Also, the Gaia source located at the coordinates of TIC 45731014 which is a red star, is more than 6 magnitudes fainter than TIC 45731009.   \\

HD 1743        	=	TIC 428415975: It was also classified as an eclipsing binary by \citet{Ijspeertetal2021} in their catalogue of OBA-type eclipsing binaries observed by {\it TESS}.  However, it blends with the nearby pulsating star TIC 428415927 which is 4 magnitudes fainter in the {\it TESS} passband. Given the low amplitude of the pulsation (0.18 mmag) this close-by star cannot be confidently excluded as their source.\\

HD 240171      	=	TIC 314034024:  The amplitudes of the pulsation frequencies are similar to that of the suspected orbital variability with a 5.1514(9) d period. The shape of the latter signal may be due to eclipses, although it appears rather reminiscent of a "heartbeat" star light curve. There are  also nearby contaminators such that we  cannot clearly identify the origin of the signals. \\

HD 277680      	=	TIC 122883754 was reported to be an eclipsing binary by \citet{Ijspeertetal2021}.
However, it is sufficiently close to TIC 122883745 (6.27 arc seconds), which is 2.3 mag fainter than the target in the {\it TESS} band such that it is not clear where the signals come from. The eclipses are shallow (40 mmag), so the 6" companion could be responsible, but the pulsations should come from the target as this companion is too cool. 
\\ 

HD 306124  = TIC 466887289: both the eclipses and the pulsational signals have very low amplitude. Hence we cannot exclude that they originate in some of the fainter close-by stars.\\

HD 46060       	=	TIC 25041731: The signals could come from more than one star as it blends with TIC 25041724 and TIC 25041738. It has a long orbital period of 20.6094(8) d. \\

HD 73903       	=	TIC 141497319:  \citet{JaschekandEgret1982} classified it as a Be Star. There are many bright nearby stars, including CPD$-$45 2774 so that it is not clear where the signals come from. \\ 

HD 76838 = TIC 30562668:  It was discovered to be eclipsing by \citet{Ijspeertetal2021} and to be a $\beta$ Cep pulsator by \citet{Chenetal2022} using {\it TESS} photometry. There are several bright early type stars around this object, two classified in SIMBAD as eclipsing binaries. Although the nearest eclipsing binary to it is a faint Gaia source, the second (CD-42 4806) 46.13 arc seconds away is only about 1 magnitude fainter. There is sufficient contamination that it is not clear whether HD 76838 is the source of the observed variability. \\

HD 92741       	=	TIC 458561474   was also identified as an eclipsing binary in the catalogue of OBA-type eclipsing binaries observed by {\it TESS} compiled by \citet{Ijspeertetal2021}.  The target has nearby stars. Examples include CPD-59 2451 and CD$-$59 3219B, which are 2 and 4.8 magnitudes fainter in the {\it TESS} band at 45 and 4.49 arc seconds respectively from it which could have possibly and significantly modulated the light curve.\\      

HD 92782       	=	TIC 458599043: There are two relatively bright stars, HD 92758 (A6/7 V) and HD 303139 (M0) of almost similar magnitudes in V band with the target in its neighbourhood (less than 100 arc seconds). It is not clear whether both the pulsations and the eclipses originate from our target of interest. The phased eclipse light curve is somewhat asymmetric.\\

Hilt 1208	=  TIC 269228628:  It was first discovered to be a $\beta$ Cep pulsator by \citet{Labadie-Bartzetal2020}. However, these authors did not observe any eclipse in its light curve. The {\it TESS} photometry shows shallow eclipses of about 38 mmag depth. Whether that eclipse originates from it or is contributed by the gaia source 88.50 arc seconds from it with a magnitude difference of 5.71164 in G band is not yet established. \\

LS VI $-04$ 19   	=	TIC 33091633: There is a nearby star TIC 33091617 that cannot be excluded to be the origin of the apparent pulsational signals. The brighter star HD 52047, 1.5 arc minute away from the target, is not.\\

NGC 2483 2  = TIC 129364361: This star is located in a dense field in an open cluster. The observed light curve is likely a superposition of the variability of several stars. \\

QX Car 	=	 TIC 469247903: It was discovered to be a spectroscopic binary by \citet{1973MmRAS..77..199T}, eclipsing by \citet{1969MNSSA..28...63C} and $\beta$ Cep candidate by \citet{SouthworthandBowman2022}. It has an orbital period of 4.47800(2) d and dominant pulsation frequency of 5.7216 c/d at a S/N of 5.5.  There is no significant contaminant in the aperture mask. Although we observed only a single $\beta$ Cep pulsation mode from it, our analysis suggests that the pulsation originates from one of the similar components \citep{1983A&A...121..271A} of the eclipsing binary.  \\

TYC 8995-5052-1	=	TIC 314139276 has an eccentric orbit, and few nearby sources that could contaminate the data. It is not clearly evident whether the signals originate from the early B-type star.  \\

V1388 Ori  = TIC 337165095:  It was discovered to be a $\beta$ Cep pulsator in an eclipsing binary by \citet{2022Obs...142..161S}.  These authors described it as a detached eclipsing binary with two early-B type components that are significantly tidally distorted. They also reported two pulsation frequencies at 2.99 and 4.0 c/d, which we confirm together with some additional lower frequency variability. If there was a $\beta$ Cephei pulsator in this system, it must be the more massive component, for which \citet{2022Obs...142..161S} derived $M=7.24$M$_{\sun}$ and $R=5.30$R$_{\sun}$. This yields a pulsation constant $Q=0.055$d, which is larger than the expected value for radial fundamental mode pulsation (see \citet{StankovandHandler2005} for a discussion). V1388 Ori could therefore be a pure g-mode pulsator.  \\ 

V446 Cep    	=	TIC 335265326 was discovered to be an eclipsing binary with an orbital period of 3.81 d by \citet{Kazarovetsetal1999} using {\it Hipparcos} satellite data and reported as a $\beta$ Cep pulsator by \citet{SouthworthandBowman2022}. Although there is a long period variable 20.30 arc seconds away, their magnitude difference  which is 6.210257 in the G band suggests that the contaminator would have at best a minimal effect on the light curve of the target. We cannot completely rule out that the pulsation frequencies in the $\beta$ Cep domain are combination frequencies of the g mode signals.  \\

V536 Mon    	=	TIC 23548300:  The binarity was first identified  by \citet{Kukarkinetal1968} and corroborated by a number of authors e.g. \citet{Kreiner2004}. It also has an eccentric orbit. Pulsations in both the $\beta$ Cep and SPB frequency regimes are reported here for the first time. Given that the star is sufficiently below the $\beta$ Cep instability strip in the theoretical HR-diagram with a mass of about $5M_\odot$ (Fig.\,\ref{fig:HR}) we are unsure whether the apparent $\beta$ pulsations are not a result of rotational splitting of the SPB pulsations or combination frequencies. This star needs further investigation. We note in passing that there is a pulsating contaminating star, TIC 23548264 ($f=7.281$\,d$^{-1}$), which is not responsible for the strongest pulsations detected for V536 Mon.\\   

V964 Sco    	=	TIC 339567760: It belongs to the cluster NGC 6231 and has  {\ensuremath{\beta}} Cep variability \citep{Meingastetal2013}. There are many possible contaminators from the cluster such that it is not certain both the eclipses and pulsations indeed come from the same star. \\

\subsection{Likely ellipsoidal variables with {\ensuremath{\beta}} Cep components\label{subsec:figures}}

BD+48 658      =	TIC 301263808: This star is an ellipsoidal variable. There is no contaminator within the aperture mask for a 5 magnitude difference. The nearby bright star BD+48 661 does also not cause notable contamination, hence, the signals come from the target.   \\ 

HD 13338       	=	TIC 347486043: It was identified as {\ensuremath{\beta}} Cep pulsator by \citet{Labadie-Bartzetal2020}. Here, we report ostensible ellipsoidal variations in addition, but cannot confidently exclude a rotational origin. There is no significant contamination. \\

HD 232874      	=	TIC 266338052: The star could be an eclipsing or ellipsoidal variable with rotational modulation \citep{Labadie-Bartzetal2020}. These authors however also classified it as a {\ensuremath{\beta}} Cep pulsator. There could be some contamination from a nearby star (TYC 3718-137-1), 31.29 arc seconds away, which is about 3 magnitude fainter. The CROWDSAP parameters are 0.87358 and 0.92231 for Sectors 19 and 59, respectively. The optical companion may be a $\delta$ Scuti star.\\ 

HD 277132      	=	TIC 121064859: It appears to be an optical double. It is most likely an ellipsoidal, less likely a rotational variable with a {\ensuremath{\beta}} Cep pulsating component.  \\  

HD 300978       =	TIC 458604640: It has a long period variable, 2MASS J10414733-5644222 (4 magnitude dimmer), 114.02 arc seconds away from it. However, our analysis suggests that the signals come from the target. \\    

HD 308106       =	TIC 465870314: No significant contamination in the photometric aperture. The star is classified as an ellipsoidal or rotational variable or eclipsing binaries with a possible {\ensuremath{\beta}} Cep component. \\

HD 327010      	=	TIC 382486122: No significant contamination in the photometric aperture. The star is classified as an ellipsoidal variable with a possible {\ensuremath{\beta}} Cep component although we cannot completely rule out a rotational origin for the 7.5537(6) d period. \\

HD 39716       	=	TIC 66975228: The star has no nearby contaminator. It is a possible ellipsoidal variable with hybrid $\beta$ Cep/SPB pulsations. \\

HD 55687       	=	TIC 178374964: There is a 1.1 mag brighter cool star (HD 55688) and two fainter hot stars all within 77" of the target. Whereas the ellipsoidal variations should come from the target, these close-by objects could contribute to the short-period variations.\\

NGC 6913 82    	=	TIC 14621767: It belongs to an open cluster NGC 6913 \citep{WangandHu2000}. It is not clear whether the long-period variation we detected is due to ellipsoidal variability or rotation. \\ 

\subsection{Rejected objects\label{subsec:reject}}

$\eta$ Ori  = TIC 4254645: It is a multiple system comprising of a detached eclipsing binary with an orbital period of 7.989 (3) d and one component showing g-mode pulsations and a non eclipsing binary with a period of 0.864 d and strong ellipsoidal variations \citep{Masonetal2001, SouthworthandBowman2022}. No $\beta$ Cep pulsation was detected.\\

BD+61 675  = TIC 84342607: The eclipses that are present in the S19 30-min data can be attributed to the nearby eclipsing binary SZ Cam (P=2.7 d). These eclipses are almost absent in the S59 2-min data, but $\beta$ Cephei pulsations are still there with the same amplitude. We conclude that BD+61 675 is a non-eclipsing $\beta$ Cephei star.\\

CD$-$59 4169  = TIC 383089500: The eclipses observed in this objects'  {\it TESS} light curve come from the  target, but the pulsations from the nearby TIC 383089586.\\

CPD$-$59 3141    	=	TIC 467066902:  The eclipses are from a nearby source and the pulsation is too weak to be localized. Both the pulsational and eclipse signals are visible in Sector 64, a likely pointer to an external origin.\\

CW Cep 	= TIC 434893323 was first discovered to be eclipsing  by \citet{Alfonso-Garzonetal2012} and as a $\beta$ Cep pulsator by \citet{LeeandHong2021}. These authors obtained  an orbital period of 2.72914(3) d and two dominant pulsation frequencies of 2.7265 and 5.3405 c/d, which are close to be harmonically related, in agreement with what we also obtained from the QLP data. As both component are located in the $\beta$ Cephei domain, it is not clear which of the components is (or whether both are) pulsating. However, the pulsation constants $Q=P\sqrt(\rho/rho_{\sun})$, with the stellar mean density derived from the parameters listed by \citet{LeeandHong2021} for those two frequencies are 0.102/0.052 or 0.110/0.056 d, respectively. We therefore conclude that the pulsations in this system are due to gravity modes. Hence, this star is disqualified as a $\beta$ Cep pulsator.\\ 

FZ CMa 	= TIC 125497512: FZ CMa was first discovered to be a double-lined spectroscopic binary by \citet{Neubauer1943}, eclipsing by \citet{MoffatandVogt1974} and $\beta$ Cep by \citet{SouthworthandBowman2022}. Detailed photometric and spectroscopic analyses of the system has been presented in literature \citep[e.g.][]{Moffatetal1983,SouthworthandBowman2022}. The pulsation spectrum is dominated by a 2.7 c/d signal and possible harmonics and subharmonics. The masses that Southworth \& Bowman derive for this system are around \textbf{$5\,M_\odot$}. All this is evidence that FZ CMa is an SPB rather than a $\beta$ Cep star. \\

HD 153772      = TIC 212415990: It was classified as $\beta$ Cep pulsator by \citet{PigulskiandPojmanski2008}. However, it is located in a very crowded field with nearby sources of relatively similar magnitude. It has very shallow eclipses as well. The light curve shape of the single short-period variation (3.276 c/d) resembles an ellipsoidal variation rather than a pulsation.\\

HD 154646      	=	TIC 380870688: The eclipses of the target are quite shallow and arise from the nearby  Gaia source 5963347209003877888. The pulsation, however, appears to emanate from HD 154646. \\

HD 217919 	=  TIC 434723918: It was discovered to be a spectroscopic binary by \citet{Garmany1972} and eclipsing $\beta$~Cep pulsator by \citet{SouthworthandBowman2022}. We obtained an orbital period of 16.2080(9) d which is in agreement with the period obtained by \citet{SouthworthandBowman2022} but in disagreement with that obtained by \citet{Garmany1972}. \citet{SouthworthandBowman2022} report significant third light and say the pulsations could arise in the third, non-eclipsing component. Again, the pulsation spectrum is consistent with a harmonic series of g mode pulsations. We suspect that this system is a rapidly oscillating representative of the SPB stars. \\

HD 308111      	=	TIC 465869053: The light curve causing the longer period variations appears to be of rotational instead of ellipsoidal origin.  \\  

LS II +23 34  = TIC 360661624: Eclipses come from target, $P_{\rm orb} = 10.9815$ d. Pulsations come from nearby LS II +23 36.\\

LS CMa = TIC 63427664:  It was classified as a poor $\beta$ Cep candidate in an eclipsing binary by \citet{SouthworthandBowman2022}.  As the single observed frequency not of orbital origin is below the $\beta$ Cep regime, we reject this object from our candidate list. \\

V Pup 	=  TIC 269562415: It was identified by \citet{Buddingetal2021} as a massive close binary system of Algol-type evolution  and a secondary that had experienced a mass ejection instead of a mass transfer.  Contrary to these authors, we do not find any coherent periodic variability apart from the eclipses. There seem to be instrumental effects and/or stochastic light variations present in the light curves of this bright object, but no $\beta$ Cephei pulsations. \\



\begin{figure}[ht!]
\plotone{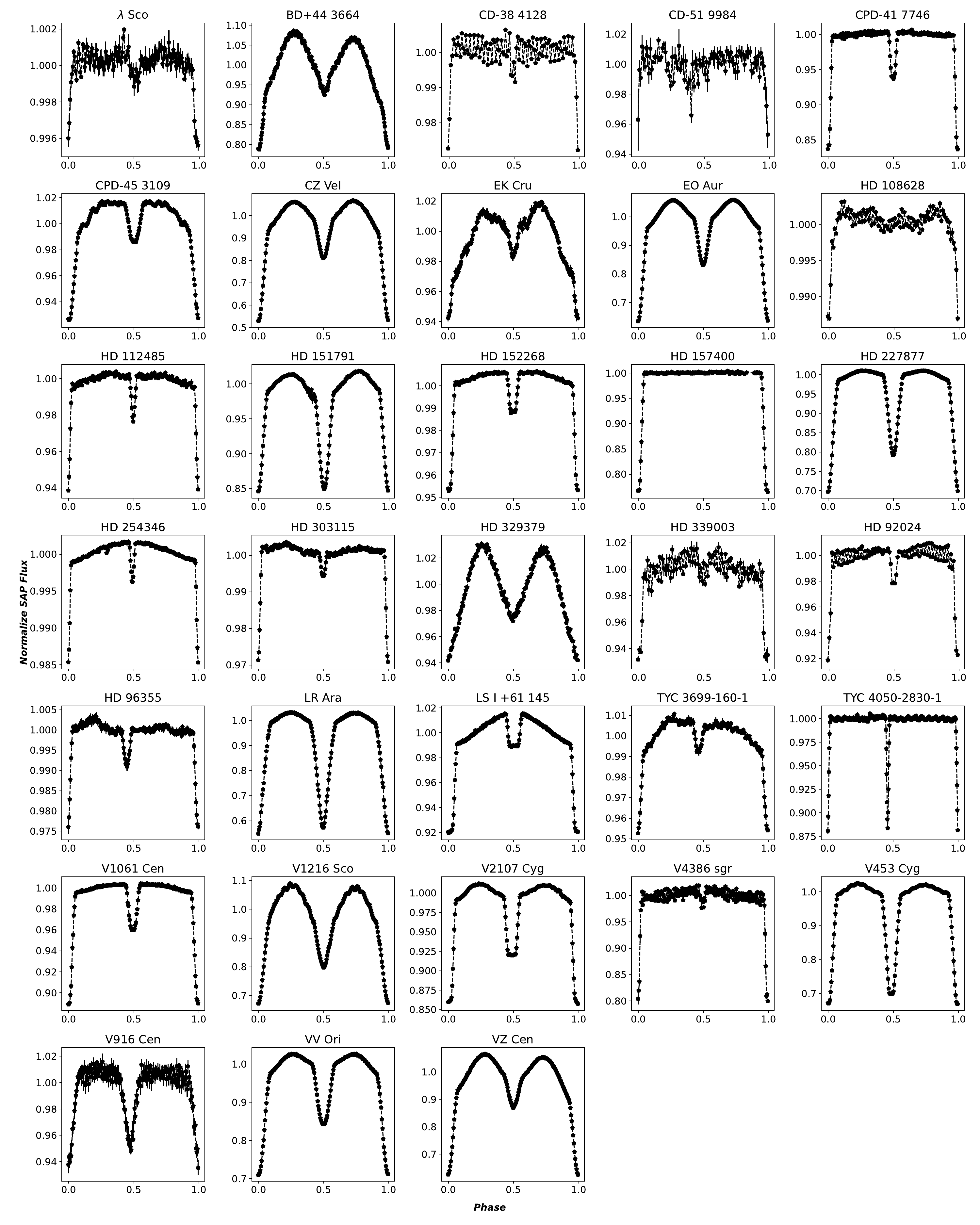}
\caption{Binned Phase diagrams of light curves of the $\beta$ Cep pulsators in eclipsing binaries after prewhitening the strongest pulsations.
\label{fig:appendix1}}
\end{figure}


\begin{figure}[ht!]
\plotone{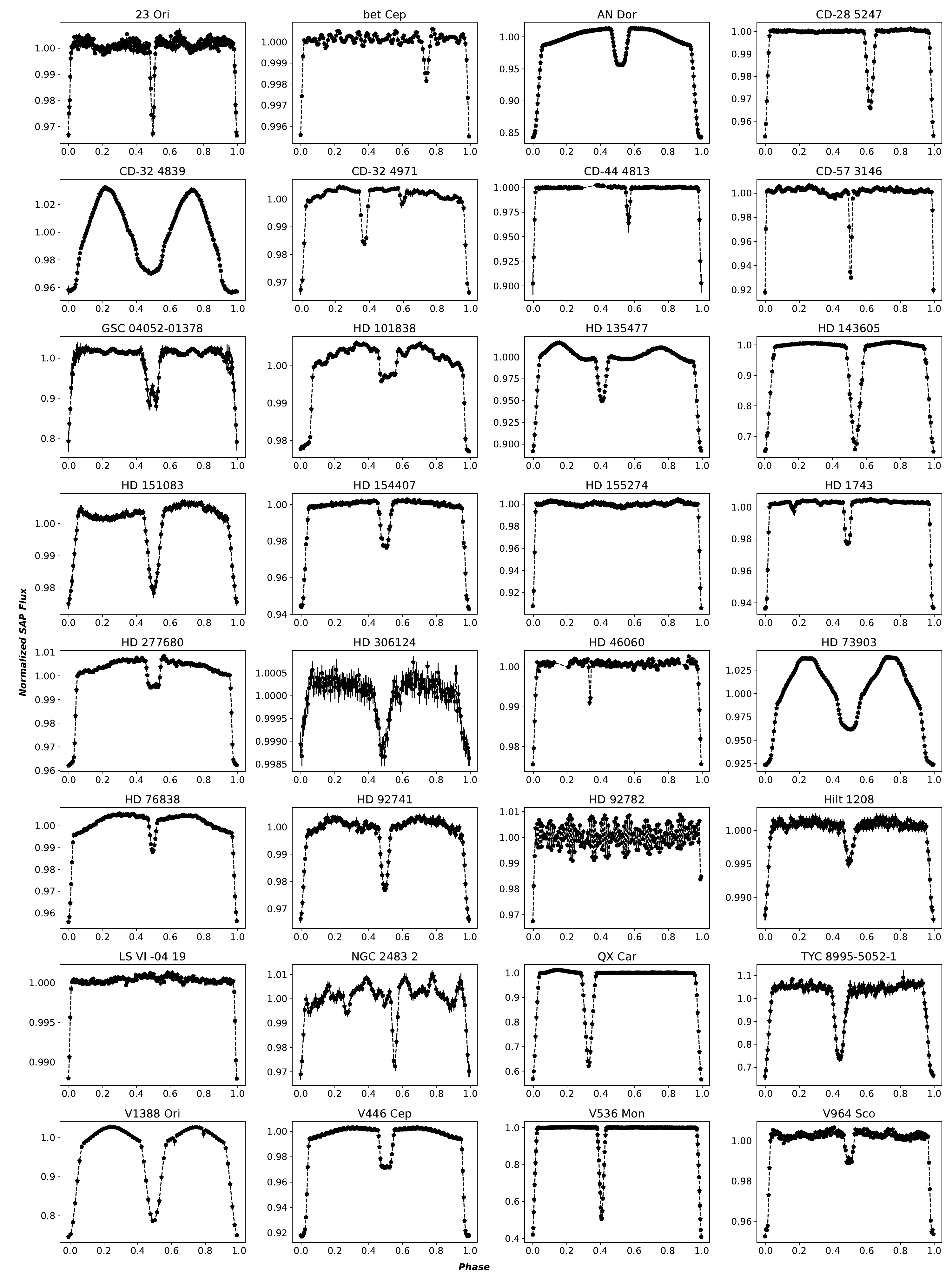}
\caption{Binned Phase diagrams of light curves of the candidates for $\beta$ Cep pulsators in eclipsing binaries after prewhitening the strongest pulsations.
\label{fig:appendix2}}
\end{figure}


\begin{figure}[ht!]
\plotone{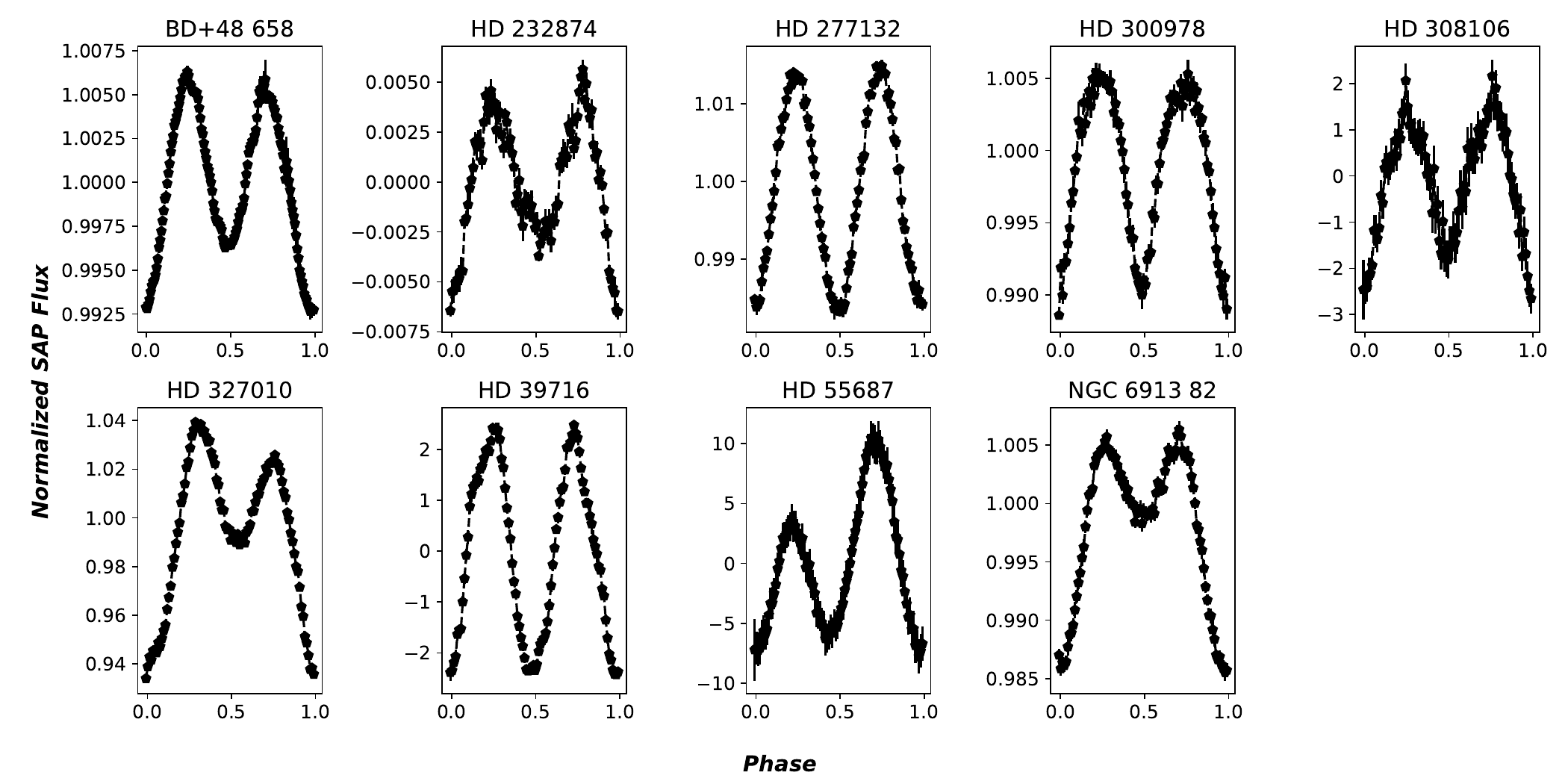}
\caption{Binned Phase diagrams of light curves of the ellipsoidal variables with $\beta$ Cep pulsating component(s) after prewhitening the strongest pulsations.
\label{fig:appendix3}}
\end{figure}



\begin{figure}[ht!]
\plotone{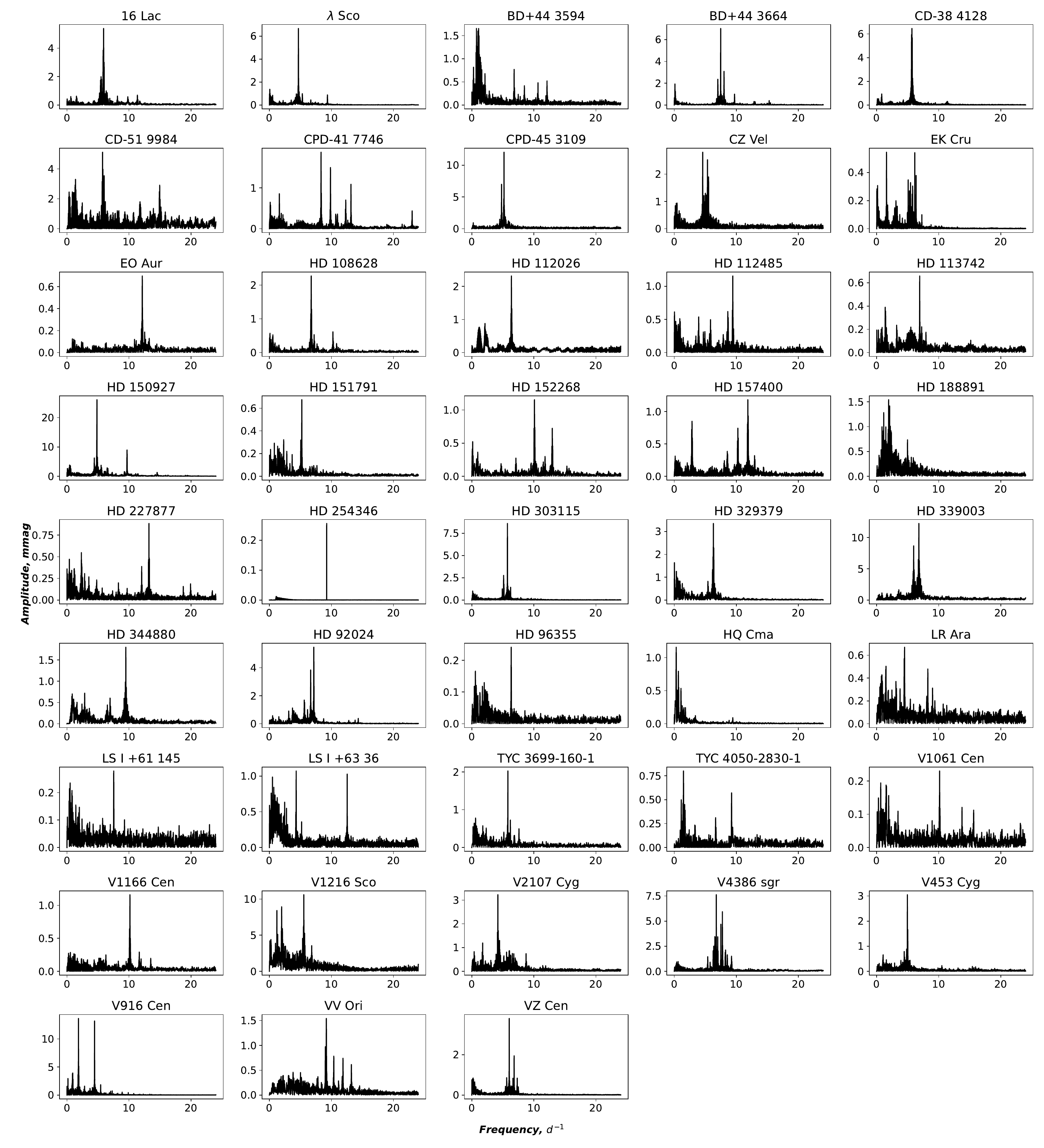}
\caption{Fourier spectra of the definite $\beta$ Cep pulsators in eclipsing binaries after removing the orbital light variations.
\label{fig:appendix4}}
\end{figure}


\begin{figure}[ht!]
\plotone{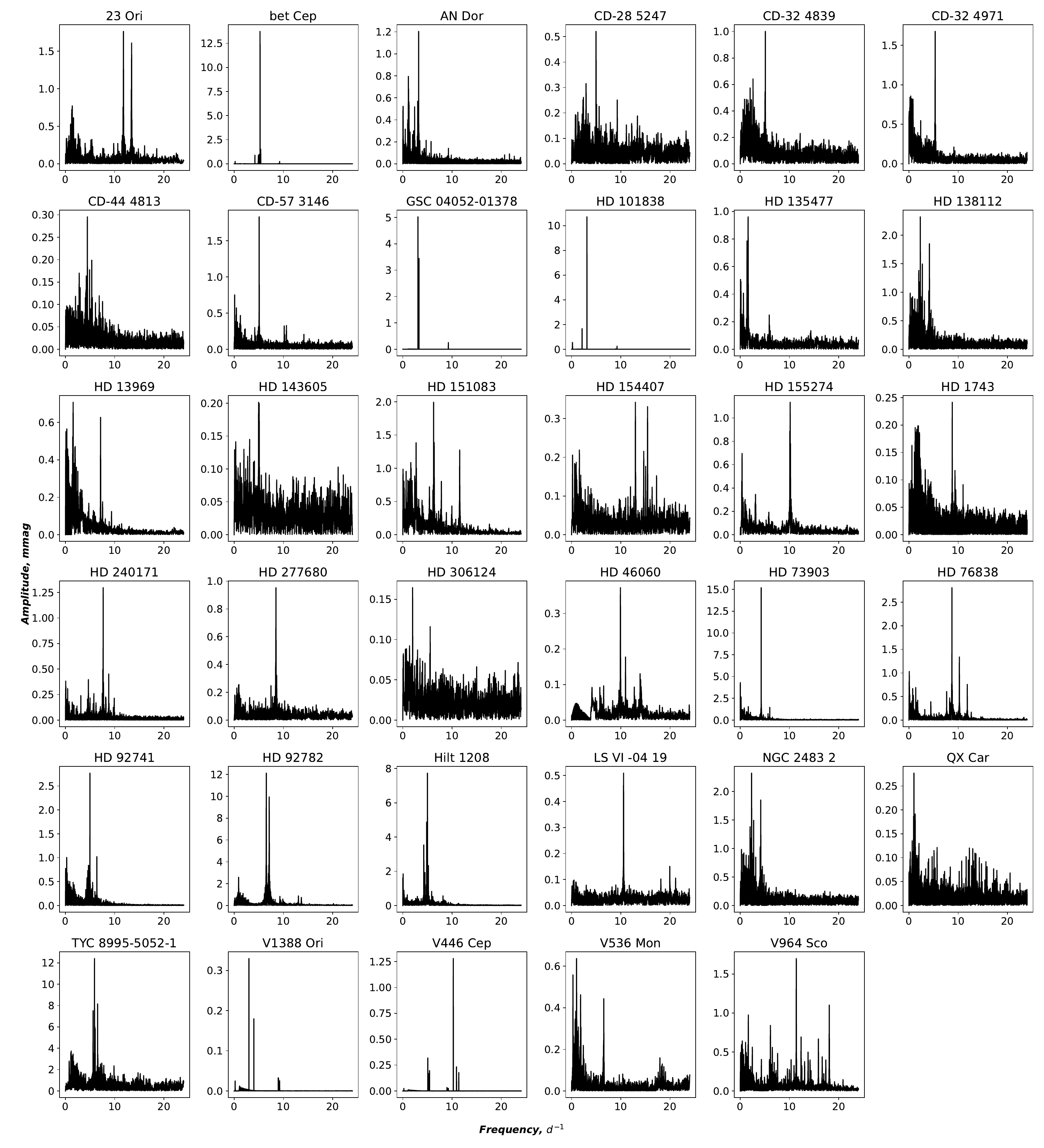}
\caption{Fourier spectra of the candidates for $\beta$ Cep pulsators in eclipsing binaries after removing the orbital light variations.
\label{fig:appendix5}}
\end{figure}


\begin{figure}[ht!]
\plotone{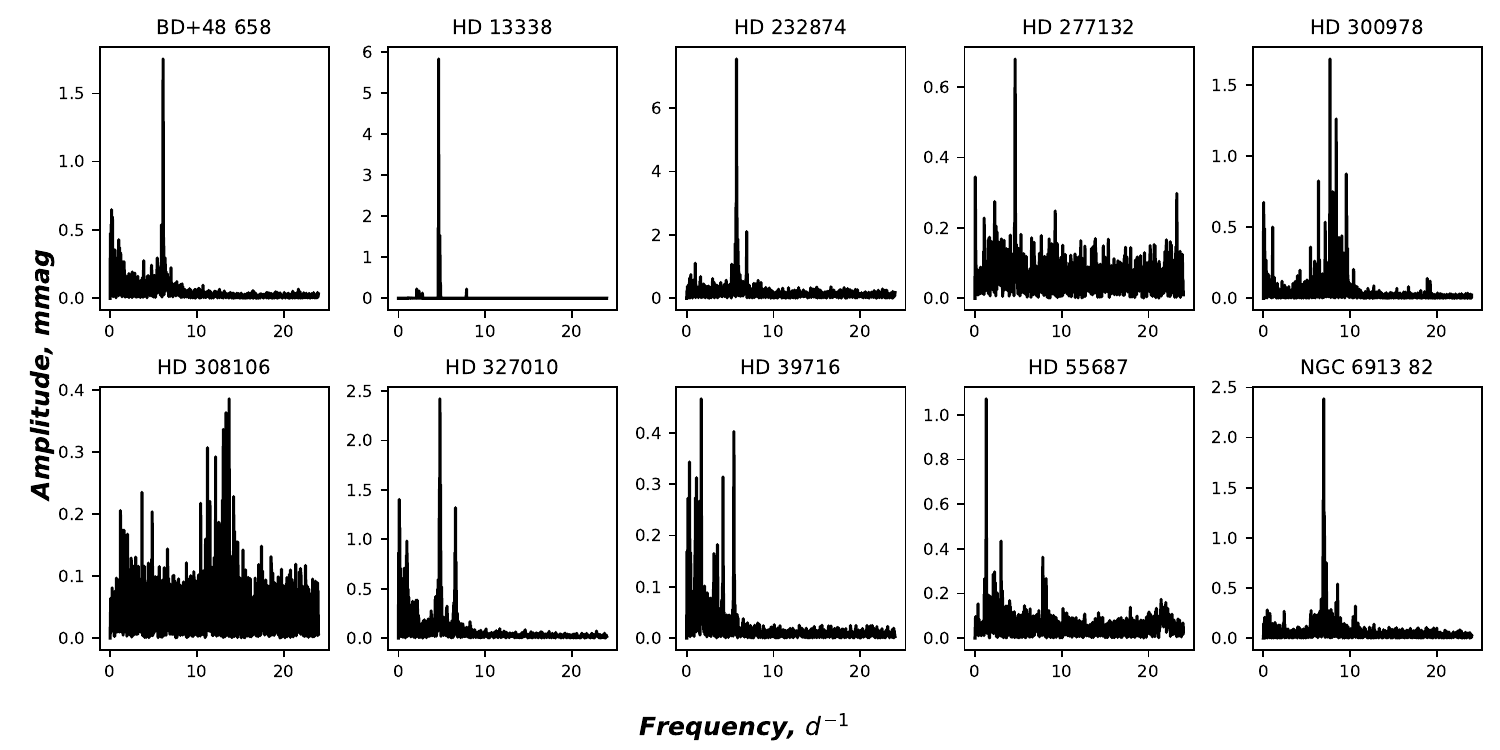}
\caption{Fourier spectra of the Likely ellipsoidal variables with {\ensuremath{\beta}} Cep components after removing the orbital light variations.
\label{fig:appendix6}}
\end{figure}

\bibliography{sample631}{}
\bibliographystyle{aasjournal}



\end{document}